\documentclass[twocolumn,aps,amsmath,amssymb,showpacs,floatfix]{revtex4}

\usepackage{graphicx}
\usepackage{dcolumn}
\usepackage{bm}

\input{epsf}
\begin{document}
\title{Quantum Computation of a Complex System: the Kicked Harper Model}

\author{B. L\'evi and B. Georgeot}

\affiliation {Laboratoire de Physique Th\'eorique, UMR 5152 du CNRS, 
Universit\'e Paul Sabatier, F-31062 Toulouse Cedex 4, France}



\begin{abstract}
The simulation of complex quantum systems on a quantum computer
is studied, taking the kicked Harper model as an example.  This
well-studied
system has a rich variety of dynamical behavior depending on parameters,
displays interesting phenomena such as fractal spectra, mixed phase space,
dynamical localization, anomalous diffusion, or partial delocalization, 
and can describe electrons in a magnetic field.  Three different 
quantum algorithms are presented and analyzed, enabling to simulate 
efficiently the evolution
operator of this system with different precision using different resources.
Depending on the parameters chosen, the system is near-integrable,
localized, or partially delocalized.
In each case we identify transport or spectral quantities which
can be obtained more efficiently on a quantum computer than on a classical one.
  In most cases, a polynomial gain compared to 
classical algorithms is obtained, which can be quadratic or less depending
on the parameter regime.  
We also present the effects of static
imperfections on the quantities selected, and show that depending on
the regime of parameters, very different behaviors are observed.
Some quantities can be obtained reliably with moderate levels of imperfection,
whereas others are exponentially sensitive to imperfection strength.
In particular, the imperfection threshold for delocalization 
becomes exponentially small in the partially delocalized regime.
Our results show that interesting behavior can be observed 
with as little as 7-8 qubits, and can be
reliably measured in presence of moderate levels of internal imperfections.

\end{abstract}
\maketitle


\section{Introduction}

In the past few years, the field of quantum information \cite{nielsen}
 has attracted
more and more attention in the scientific community.
Among the most fascinating promises of this domain is the possibility
of building a quantum computer.  Such a quantum processor can use
the superposition principle and the interferences of quantum mechanics
to perform new types of algorithms which can be much more efficient
than classical algorithms.  Celebrated examples
are Shor's algorithm which factors large integers exponentially faster than
any known classical algorithm \cite{shor}
and Grover's algorithm which searches unstructured lists quadratically faster
than classical methods \cite{grover}.  Another type of quantum algorithms 
concerns the simulation of physical systems. Examples include
many-body quantum systems \cite{lloyd}, 
classical and quantum spin systems \cite{spin}, classical dynamical systems
\cite{cat,recurrence}.  Algorithms implementing
quantum maps are especially interesting, since the systems simulated 
have simple equations of motion but can display very complex behaviors.
Their simplicity enables to simulate them with a small number of qubits.
For example, it is possible to simulate efficiently the baker map 
\cite{schack} (experimental implementation with the NMR technique 
has already been performed \cite{baker}), the quantum kicked rotator 
\cite{GS,Levi}, the sawtooth map \cite{complex}, or
the tent map \cite{frahm}.  In such algorithms, it is important 
to determine which physical quantities can be obtained accurately
through measurement on the quantum computer, and what is the total
algorithmic complexity of the whole process.  It is equally important
to determine the effects of errors in the computation to assess
the efficiency of the algorithm on a realistic quantum computer.

In the present paper, we will study in detail an important example of quantum 
map, namely the kicked Harper model.  The Hamiltonian of this system 
has a simple form,
yet displays many interesting physical features not present in 
quantum maps previously
studied in this context, such as fractal spectra, stochastic web, 
anomalous diffusion, or coexistence of localized and delocalized states.
It was introduced in the context of solid state
physics (motion of electrons in presence of magnetic field), and has
been the subject of many studies.  
Using this model as a test ground, we will present three different ways
of simulating the quantum map on a quantum computer, two of them
inspired by previous works, and compare
their efficiency.  We will then present examples of physical quantities 
which can be obtained on a quantum computer.  It turns out that
depending on the parameters of the system, at least polynomial speed-up
compared to classical algorithms can be obtained for different quantities.
Numerical simulations and analytical estimations 
will also evaluate the effects of imperfections in the quantum
computer on the estimation
of these quantities.  

\section{Harper and kicked Harper models}

The Harper model was introduced in 1955 \cite{harper}
to describe the motion of electrons in 
a two-dimensional lattice in presence of a magnetic field.
Its Hamiltonian reads 

\begin{equation}
\label{harper}
H_0(I,\theta)=\cos(I) +\cos(\theta)
\end{equation}

This Hamiltonian has been the subject of many studies 
(see for example \cite{hofstadter,jeanbel,geisel1,wilkinson,
pichard}), 
but its dynamics is somewhat restricted by the fact that it 
describes an integrable system.  A generalization of this model 
has been introduced some time ago; it is called the kicked 
 Harper model:
\begin{equation}
\label{kharper}
H(I,\theta,t)=L\cos(I) +K\cos(\theta)\sum_{m} \delta(t-m),
\end{equation}
where $m$ runs through all integers values and $K, L$ are constants.
This Hamiltonian reduces to (\ref{harper}) in the limit $K=L \rightarrow 0$,
but has a more complex dynamics depending on the parameters.  Its dynamics
between two kicks can be integrated to yield the map:
 
\begin{equation}
\label{map}
\left\{
\begin{array}{rcl}
\bar{I}      & = & I + K \sin{\theta} \\
\bar{\theta} & = & \theta - L \sin{  \bar{I}}
\end{array}
\right.
\end{equation}

As in the case of the kicked rotator, there is a classical periodicity
in both $\theta$ and $I$.  Thus the phase space is composed of cells
of size $2\pi \times 2\pi$ where the same structures repeat themselves.

This map (\ref{map}) 
has been related to the motion of electrons in a perpendicular 
magnetic and electric fields, and also to the problem of stochastic heating 
of a plasma in a magnetic field.

\begin{figure}
\includegraphics[width=\columnwidth]{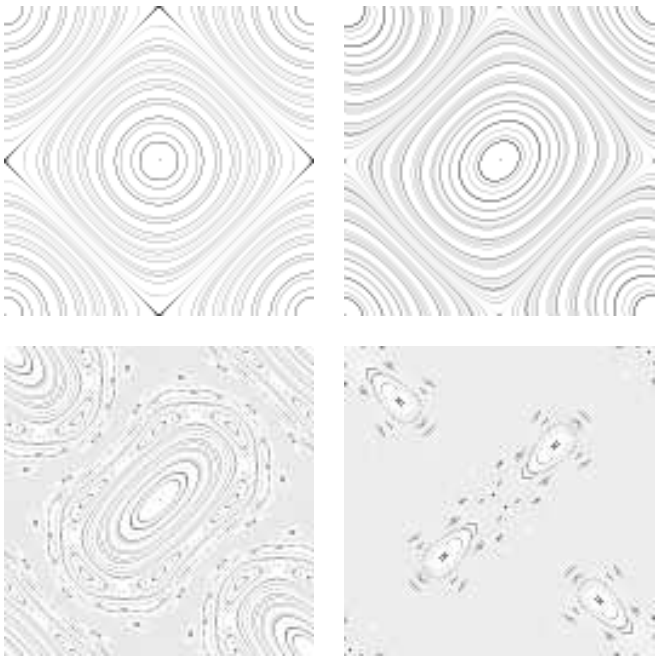}
\caption{\label{phases_harper} Phase space of the classical kicked
Harper model: $K=L \rightarrow 0$ (Harper model) (upper left), $K=L=0.5$
(upper right), $K=L=1.5$ (lower left), $K=L=2.5$ (lower right)
(10000 iterations of 256 classical orbits). One cell of size
$2\pi \times 2\pi$ is shown, the phase space being periodic.}
\end{figure}

The quantization of (\ref{kharper}) yields a periodic Hamiltonian which after
integration over one period yields a
unitary evolution operator
acting on the wave function $\psi$
\begin{eqnarray} 
\label{qharper}
\bar{\psi} = \hat{U} \psi = e^{-iL\cos(\hbar\hat{n})/\hbar} 
e^{-iK\cos(\hat{\theta})/\hbar} \psi,
\end{eqnarray}
where $\hat{n}=-iQ  \partial / \partial \theta $ 
and $\psi(\theta+2Q \pi )=\psi(\theta)$.

This system has been the subject of many studies in the past few years,
which focused on localization properties 
\cite{leboeuf,lima,artuso1,artuso,borgonovi,dana,geisel,prosen}, 
tunneling
properties \cite{mouchet,ullmo}, etc...

\begin{figure}
\includegraphics[width=\columnwidth]{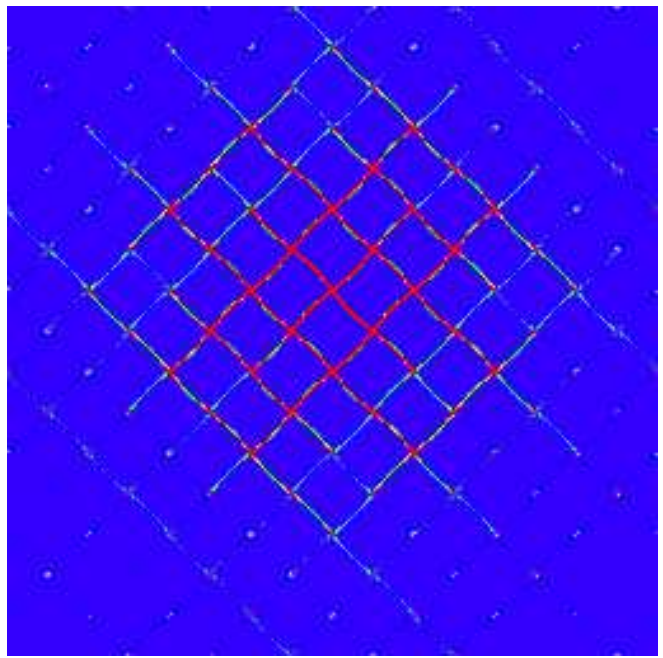}
\caption{\label{stochastic} (Color online)
Example of stochastic web in the kicked
Harper model.  Here $K=L=0.5$, phase space is $8\times 8$ cells of size
$2\pi \times 2\pi$, the figure shows positions after $t=1000$ iterations
of $10^6$ classical trajectories initially distributed
according to a gaussian centered half a cell
above the center with standard deviation 
$\sqrt{2\pi/2^{25}}\approx 0.0004$. 
Color (grayness) shows density of points, from
red (gray) (maximal value) to blue (black) (minimal value).}
\end{figure}

\begin{figure}
\includegraphics[width=\columnwidth]{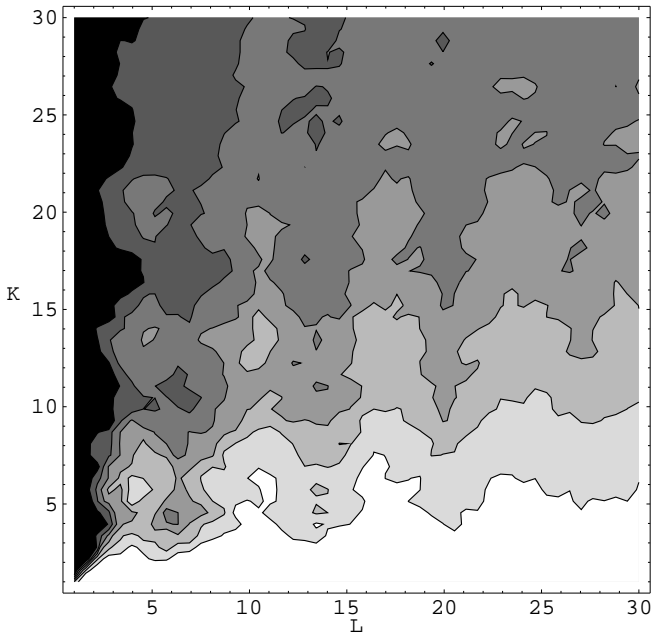}
\caption{\label{carte} Map of delocalization in the $(K,L)$ plane.
Grayness represents the Inverse Participation Ratio 
$\xi=1/\Sigma_n |\psi(n)|^4$ (IPR), a measure of delocalization of states,
from $\xi=1$ (state localized on one momentum state) to $\xi=N_H$ 
(totally delocalized state)($N_H$ is the dimension of the Hilbert space). 
Contour lines correspond to values of $\xi$ ranging from $32$ to $192$
by increments of $32$,
$N_H=2^9$, 
$\hbar/2\pi=(13-\sqrt{5})/82$ (actual value is the nearest fraction with
denominator $2^{9}$). White corresponds to 
lowest values, black to maximal values of $\xi$. Each $\xi$ value is obtained
by averaging over all eigenstates of the evolution operator $\hat{U}$
of (\ref{qharper}).
} 
\end{figure}

In the limit $K=L \rightarrow 0$ 
the system is classically integrable. For small
$K,L$, chaos begins to appear around separatrices, and spreads
over larger and larger phase space areas as $K,L$ increase
(see Fig.\ref{phases_harper}).  In the regime of small $K,L$,
classical transport from cell to cell is possible only in the very
small chaotic zones around separatrices.  For $K=L$, 
this network of thin chaotic zones
surrounding large islands is called ``stochastic web''
(see Fig.\ref{stochastic}). For intermediate
values of $K,L$, the phase space is mixed,
with integrable islands separated by large chaotic zones.
 For larger $K,L$, classical chaos is present
 in most of the phase space (cf Fig.\ref{phases_harper}), and a typical
trajectory will diffuse through the system.
The quantum dynamics is related to these classical
properties, but shows some striking differences.
In the limit $K=L \rightarrow 0$, 
the system is integrable, wave functions
are concentrated around classical tori, but complexity manifests itself
in the spectrum of the Hamiltonian, which is fractal
(``Hofstadter butterfly'').
   For small $K,L$
the motion of a quantum wave packet is dominated by the presence
of classical invariant curves; the wave packet can spread 
in between these curves, or cross them by quantum tunneling.
 For larger $K,L$, in the regime
of classical diffusion, as in the
kicked rotator, a phenomenon similar to Anderson localization of electrons in
disordered solids takes place.  Through this phenomenon, called 
dynamical localization, a wave packet started at some value of momentum $n$ 
will first spread, but contrary to classical trajectories this spreading
will saturate.  This corresponds to the fact that eigenfunctions $\psi_a (n)$
of $\hat{U}$ in (\ref{qharper}) in momentum space (they are called 
Floquet eigenfunctions since they correspond to the action of the evolution
operator during one period) are
exponentially localized.
Their envelopes obey the law  $\psi_a (n) \sim \exp(-|n-m|/l)/\sqrt{l}$ 
where $m$ marks 
the center of the eigenstate and $l$ is the localization length.
This phenomenon is especially visible for moderate values of $K$, where
all eigenfunctions are localized.  For larger values of $K$, the system 
undergoes a transition: some eigenfunctions are still localized, but
more and more are delocalized (ergodic) and spread over the whole system.
This coexistence of localized and delocalized states gives rise to
specific physical properties.  Indeed, it
is very different from what happens in the kicked rotator model,
where usually all states are localized once classical chaos is present (see
for example \cite{Levi}) or in the Anderson transition (investigated 
in \cite{pomerans}) where the transition separates a regime where
all eigenstates are localized from a regime where all are delocalized.
In this regime of partial delocalization,
an initial wave packet
will spread, but a certain fraction of the total probability will
remain localized.  In addition, the diffusion of probability in
momentum space has been shown numerically to be anomalous,
with an exponent depending on the parameter values 
\cite{geisel1,artuso1,artuso}.
These properties are summarized by the phase 
diagram of Fig.\ref{carte}.  
 Different quantities can be obtained in these
different regimes with the help of a quantum computer. 

The phase space can be decomposed in cells of size $2\pi \times 2\pi$.
Its global topology depends on boundary conditions.
For a system of size $N_H$, if the phase space is closed with
periodic boundary conditions, with
respectively $Q$ and $P$ cells
in the $\theta$ and $n$ directions, then $\hbar=2\pi PQ/N_H$.
Therefore if one wants to keep $\hbar$ constant, 
the product $PQ$ should be chosen such that
$PQ/N_H$ is the closest rational to $\hbar/(2\pi)$.
For most of the results of this paper,
the phase space will be a cylinder closed in the $\theta$ direction ($Q=1$)
and extended in the direction of momentum,
and transport properties will be studied in the momentum direction, as in
the kicked rotator.  In this case $\hbar/(2\pi)$ was set to 
$1/(6+1/((\sqrt{5}-1)/2))=(13-\sqrt{5})/82$ as in \cite{artuso} to avoid 
unwanted arithmetical effects.  The choice of a constant $\hbar$ 
implies that changing the number of qubits leads to increasing 
the size of phase space (number of cells) in the $n$ direction.
Only for the study of the stochastic web 
present at small $K=L$ (subsection IV A)
will the phase space be extended in both directions
and its size (number of cells) fixed.  In this case increasing the number of
qubits leads to smaller and smaller $\hbar$.


\section{Simulating the time evolution: three possible algorithms}

The evolution operator (\ref{qharper}) is composed of two
transformations which are diagonal in respectively the momentum
and position representations.  This form is general for
a family of kicked maps such as the kicked rotator, sawtooth map,
and others.  On a classical computer, the fastest way to implement 
such an evolution operator on a wave function of $N_H$ components is
to use the Fast Fourier Transform algorithm to shift back and forth
between the $n$ and $\theta$ representations, and
to implement each operator by direct multiplication in the basis
where it is diagonal.  In this way, $O(N_H\log N_H)$ classical operations are 
needed to implement (\ref{qharper}) on a $N_H$-dimensional vector. 
On a quantum computer, it is possible to use the Quantum Fourier Transform 
(QFT) to shift between momentum and position representations, using
$O((\log N_H)^2)$ quantum gates.  In each representation, one has
then to implement the multiplication by a phase, 
$e^{-iL\cos(\hbar\hat{n})/\hbar}$ and 
$e^{-iK\cos(\hat{\theta})/\hbar}$.  

In the following we will envision three different strategies
to implement these diagonal operators:
\begin{itemize}
  \item exact computation using additional registers to hold the values
        of the cosines
  \item decomposition into a
    sequence of simpler operators which are good approximations
    during short time-slices 
  \item direct computation,
   the cosine function being approximated by 
    a (Chebyshev) polynomial
\end{itemize}

  The first one 
is in principle exact, but requires extra registers, and 
was already proposed in \cite{GS}.
The second one has some similarities with 
the one explained in \cite{pomerans}
for another system. The third one was not used in the context
of quantum algorithms to the best of our knowledge,
although the method is well known in computer science
(see for example \cite{dobrovitski} for a recent use 
of this method to simulate many-spin systems on a classical
computer).
We note that an approximate algorithm to
simulate the kicked Harper for long
time was used in \cite{saraceno}; however, in that paper the
aim of the authors was different, since
they only wanted to construct efficiently a good approximation
of the ground
state wave function in order to use it for generating phase space
distributions of other systems, and it is not clear that the method works
for other purposes.
We also note that the simulation of the Harper model on optical lattices
was envisioned in \cite{zoller}.  In the following discussions, 
we note by $n_q$ the total number of qubits including ancilla and 
workspace qubits, and 
$N=2^{n_q}$ is the total dimension of the Hilbert 
space of the quantum computer.
We note $n_r$ with 
$n_r\leq n_q$ the number
of qubits describing the Hilbert space of the kicked Harper model
(i.e. the wave function evolved through (\ref{qharper}) is $N_H$-dimensional
with $N_H=2^{n_r})$, and $n_g$ is the number of elementary quantum gates
used for one iteration of the quantum map (\ref{qharper}).

\subsection{Exact algorithm}

This approach is similar to the one taken in \cite{GS} for
the quantum simulation of the kicked rotator.
In each representation, 
the value of the cosines is built on a separate register, and
then transferred to the phase of the wave function by
appropriate gates. 

If one starts with a $N_H$-dimensional wave function 
$|\psi\rangle = \sum_{i=0}^{N_H-1} a_i |\theta_i\rangle$ in the $\theta$
representation, with $N_H=2^{n_r}$, then the first step is to perform:
$$ \sum_{i=0}^{N_H-1} a_i |\theta_i\rangle|0\rangle \to
\sum_{i=0}^{N_H-1} a_i |\theta_i\rangle|\cos\theta_i\rangle $$

To this aim, 
the $2n_r$ values $\cos (2\pi/2^j)$ and $\sin (2\pi/2^j)$,
for $j=1,..,n_r$ are first precomputed classically
with precision $2^{-n_p}$ with for example $n_p =2n_r$
using a recursive method based
on Moivre's formula; then since
$ \theta_i =\sum_{j=1}^{n_r} \beta_{ij} 2\pi/2^{j}
\mbox{ with } \beta_{ij}=0 \mbox{ or } 1$,
one has:

\begin{eqnarray}
\exp(i\theta_i) & = & \prod_{j=1}^{n_r} \exp(i\beta_{ij} 2\pi/2^{j}) \nonumber \\
& = & \prod_{j=1}^{n_r} (\cos(\beta_{ij} 2\pi/2^{j}) 
+i \sin(\beta_{ij} 2\pi/2^{j})) \nonumber
\end{eqnarray}

This enables to compute $|\cos\theta_i\rangle$ for each $\theta_i$ 
in $n_r$ multiplications by $\exp(i 2\pi/2^{j})$ 
conditioned by the values of $\beta_{ij}$, 
needing in total $O(n_r^3)$ quantum gates.

Then once the binary decomposition of $\cos\theta_i$ is present
on the second register, conditional application of the
$n_r$ one-qubit gates $\left( \begin{array}{cc} 1&0 \\
0&\exp (-iK2^{-j}/\hbar) \end{array} \right)$ yields the
state:

$$\sum_{i=0}^{N_H-1} a_i \exp(-iK\cos(\theta_i)/\hbar)|\theta_i\rangle|\cos \theta_i \rangle$$

Then the cosines in the last register are erased by running
backward the sequence of gates that constructed them, and one ends up
with the state 
$\sum_{i=0}^{N_H-1} a_i \exp(-i K\cos(\theta_i)/\hbar   )|\theta_i\rangle|0\rangle$,
which is the result of
 the action of the unitary operator $\exp(-iK\cos(\hat{\theta})/\hbar)$
on $|\psi\rangle$.

Then the use of the QFT which needs $O(n_r^2)$ quantum gates shifts
the wave function to the momentum representation, and exactly the same
technique as above enables to implement the operator 
$\exp (-iL\cos(\hbar\hat{n})/\hbar)$ in $O(n_r^3)$ quantum gates.
A second QFT enables to go back to the $\theta$ representation.

The whole process implements one iteration of the evolution operator
$\hat{U}$ in $O(n_r^3)$ operations, with exponential precision.
This algorithm is therefore efficient, and precision can be increased 
exponentially at a cost of polynomial number of operations.
On the other hand, the drawback of this approach is the need of several 
extra registers (one holding the values of the cosines, 
plus others for the workspace of the computation) and a relatively large 
number of gates.  In the present status of experimental implementations
of quantum computers, both the number of qubits and the number of gates
that can be applied are very expensive resources.
In the following, we will therefore expose two alternative strategies
to implement $\hat{U}$, which are much more economical in the use of resources,
but involve additional approximations.

\subsection{Slice method}
\label{sec:slices}

This technique enables to compute the operator $\hat{U}$ of (\ref{qharper})
without explicitly
calculating the cosines.  It approximates $\hat{U}$ by a sequence of
many operators, each of them being easier to compute.  It can be viewed as
``slicing'' the operator into elementary operators.

As above, we start with a $N_H$-dimensional wave function 
$|\psi\rangle = \sum_{i=0}^{N_H-1} a_i |\theta_i\rangle$ in the $\theta$
representation, with $N_H=2^{n_r}$. In general, suppose
we want to perform the operator
$$ U_k=e^{-ik\cos{(p\,\hat{\theta}})} $$

In the $\theta$ representation, this operator is diagonal, so we just have to
multiply each state by the phase $ e^{-ik\cos{(p\,\theta)}} $.
First, we write $\theta$ as
\begin{equation}\label{eq:thetadef}
\theta=\frac{2\pi}{N_H}\sum_{i=0}^{n_r-1} d_{i}2^{i}
\end{equation}
where the $d_i$'s are the binary expansion of $\theta$ and $N_H=2^{n_r}$. 
If $p=2^{a}m$ with $m$ odd, then
$$p\,\theta=\frac{2\pi m}{N_H}\left(\sum_{i=0}^{n_r-a-1} d_{i}2^{i+a}\right) \bmod 2\pi$$
Thus $U_k$ is equivalent to applying $$ e^{-ik\cos{(m\,\theta)}} $$ on the $n_r-a$ first qubits.
In the following, we will suppose that $p$ is odd ($a=0$) for the sake of simplicity.\\

\begin{figure*}
\includegraphics[width=\textwidth]{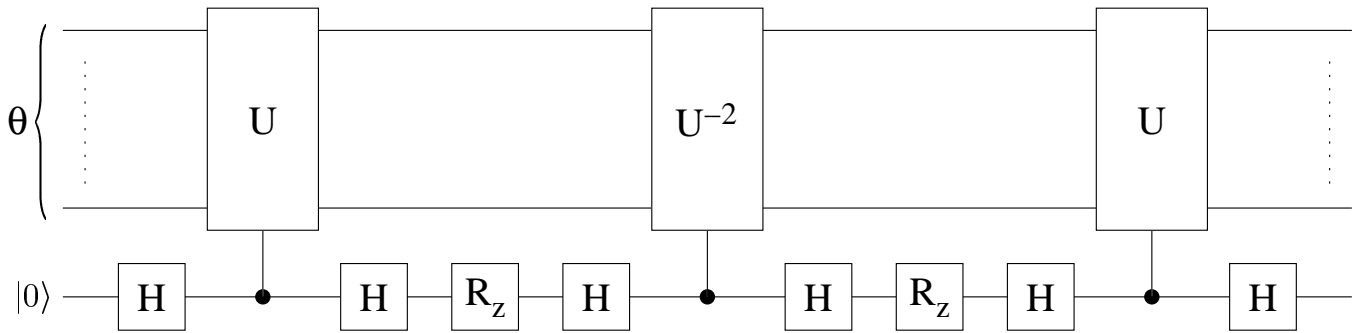}
\caption{\label{slices} Gate sequence for slices algorithm. $R_z$ are Z
rotations of angle $-\alpha$.}
\end{figure*}

With the help of one ancilla qubit, let us perform the following sequence,
where all gates are applied to the ancilla (initially set to $|0\rangle$),
except for $C_U$ which is the operator $U$ applied on the principal register, 
controlled by the ancilla 
(the gate sequence is also displayed on Fig.\ref{slices}):
$$M(\alpha,U) = H C_U H e^{i\frac{\alpha}{2}\sigma_z} H C_{U^{-2}} H e^{i\frac{\alpha}{2}\sigma_z} H C_U H$$
This product is equal to
\begin{eqnarray*}
M(\alpha,U) & = & \cos^2{\frac{\alpha}{2}} -\sin^2{\frac{\alpha}{2}}\frac{U^2+U^{-2}}{2} \\
            &   & +i\sin{\alpha}\frac{U+U^{-1}}{2}\sigma_z -i\sin^2{\frac{\alpha}{2}}\frac{U^2-U^{-2}}{2i}\sigma_x \\
            & = & 1 +i\alpha\frac{U+U^{-1}}{2}\sigma_z +O(\alpha^2) \mbox{ for } \alpha \ll 1
\end{eqnarray*}
If we take $U=e^{ip\theta}$
$$M(\alpha,U) = 1 +i\alpha\cos{(p\,\theta)}\sigma_z +O(\alpha^2)$$
since the ancilla qubit is in the $|0\rangle$ state,
$$M(\alpha,U) \approx e^{i\alpha \cos{(p\,\hat{\theta}})}$$ 
The kick operator can then be performed by $n_s\gg1$ applications of $M(\alpha,U)$
$$U_k \approx {M(\alpha,U)}^{n_s} \mbox{ with } \alpha=\frac{-k}{n_s}$$

A more accurate expansion can be obtained by symmetrizing $M(\alpha,U)$
\begin{eqnarray*}
\widetilde{M}(\alpha,U) & = & M\left(\frac{\alpha}{2},U\right)M\left(\frac{\alpha}{2},U^{-1}\right) \\
                    & = & 1 +i\alpha\frac{U+U^{-1}}{2}\sigma_z 
                            -\frac{\alpha^2}{2}{\left(\frac{U+U^{-1}}{2}\right)}^2 +O(\alpha^3)
\end{eqnarray*}
Thus $U_k \approx {\widetilde{M}(\alpha,U)}^{n_s}$ up to order $2$ in $\alpha$.

In this way, once a certain precision has been fixed, $n_s$ can be chosen such 
that the error is small enough.  

If we apply this strategy to the kicked Harper model, the method is therefore
to first compute $\exp(-iK\cos(\hat{\theta})/\hbar)$ through the technique
above ($k=K/\hbar$, $p=1$), then use a QFT to shift to the momentum
representation.  In the $n$ representation, the operator
$\exp (-iL\cos(\hbar\hat{n})/\hbar)$ can be cast in the form above for 
$\hbar=2\pi m/N_H$, with  $p\rightarrow m$, $k \rightarrow L/\hbar$ and
$\theta \rightarrow 2\pi n/N_H$.
The use of a QFT then shifts back the wave function to
the $\theta $ representation.  

The evolution of a $N_H$-dimensional wave function with $N_H=2^{n_r}$
through one time-slice
is efficient, costing $O(\log N_H)$ quantum operations.  Indeed,
for $n_s$ slices, one diagonal operator in (\ref{qharper})
is implemented
in $4+2(n_r-a) +(n_s-1) (7+2(n_r-a))$ elementary gates, with $a \leq n_r$.
The number of slices fixes the precision.  If one requires a fixed
precision, independent of the number of qubits, then the whole method
is efficient, iterating $\hat{U}$ in $O((\log N_H)^2)$ operations (the most 
costly operation asymptotically being the QFT).  
However, if one requires the precision
to increase with $N_H$, then the method becomes less efficient.  
This algorithm is quite economical
in qubits, since to simulate a wave function on a Hilbert space of
dimension $2^{n_r}$, only $n_q=n_r+1$ qubits are needed.  One should note
that for large number of slices, their computation dominate the computation
time although asymptotically the QFT dominates.  In all numerical simulations
we performed, the slice contribution was indeed dominant.

\begin{figure}
\includegraphics[width=\columnwidth]{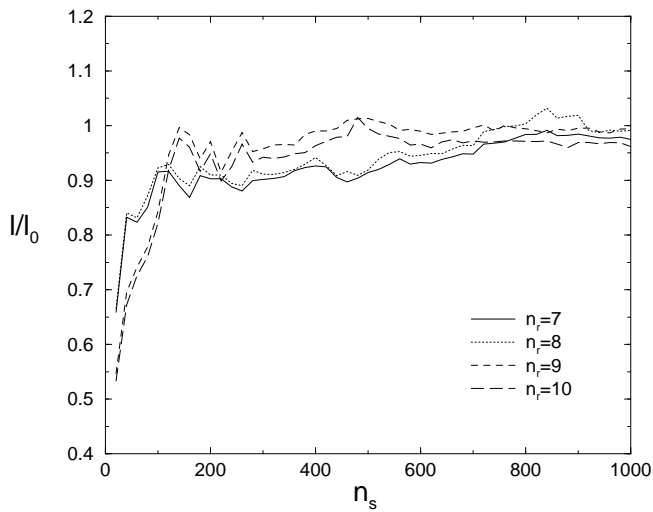}
\caption{\label{longueur_slices}
Localization length computed with the slice method
over exact localization length $l_0$
as a function of number
of slices per iteration. Localization length
is extracted after $t=1000$ iterations.
Initial state is $|\psi_0\rangle=|0\rangle$, with $K=1,L=5$, 
$\hbar/2\pi=(13-\sqrt{5})/82$ (actual value is the nearest fraction with
denominator $2^{n_r}$ with $n_r=n_q-1$).}
\end{figure}

To precise the accuracy of the method, we show examples of
the localization length
in the localized regime
as a function of number of gates in Fig.\ref{longueur_slices}.  The convergence
with increasing number of slices (gates) is clearly seen, although
for small number of gates oscillations are present.  Data from $n_r=7,8$
and $n_r=9,10$ are close to each other due to the structure of the algorithm~:
indeed, $\hbar/(2\pi)$ is approximated by its closer approximants $m/N_H$, and 
incrementing $n_r$ by one 
changes every other time the value
of $\hbar$.
No major modification is seen
in the numerical data for increasing $n_r$, indicating that in this regime
$n_s$ does not need to be drastically changed with $n_r$.

\begin{figure}
\includegraphics[width=\columnwidth]{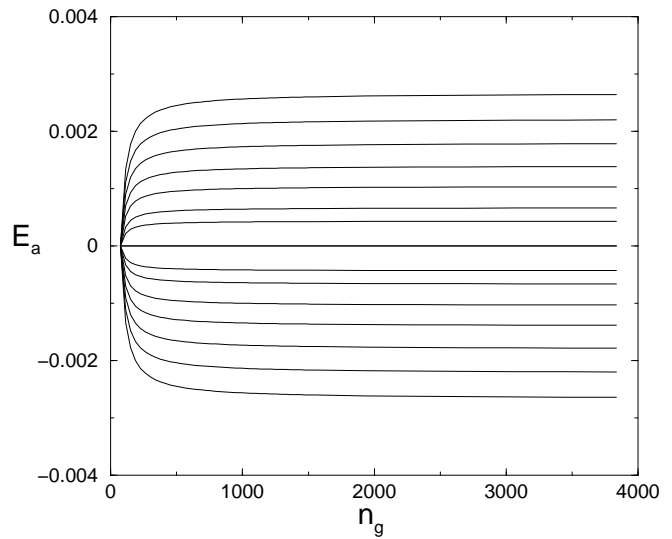}
\caption{\label{slices_spectre_portes} Eigenphases of (\ref{qharper})
 as a function of number
of gates with the slice method: only $16$ values
are shown. Here $n_r=6$ ($n_q=n_r+1$), $\hbar=2\pi/2^6$, 
$K=L=10^{-3}$}. 
\end{figure}

One may think that the spectrum is
a much more sensitive quantity than the localization length.
In Fig.\ref{slices_spectre_portes}, we
display the convergence for the spectrum of $\hat{U}$ for
$K$ and $L$ small, in a parameter
regime close to the fractal ``butterfly''
visible for the unkicked Harper model (see Fig.\ref{papillon8}). 
The quantities displayed
correspond to eigenphases $E_a$ where 
$\hat{U}|\psi_a\rangle =\exp ( iE_a) |\psi_a\rangle$ for some $|\psi_a\rangle$.
The matrix of the operator $\hat{U}$ of (\ref{qharper}) 
is built by evolving through the slice 
method explained above the basis vectors, and then diagonalized.
Convergence can be achieved with
a few hundred time-slices.
  Due to numerical limitations, we cannot 
present data for different values of $n_r$, but we do not expect 
any drastic modification.

In the subsequent sections, numerical simulations
of this algorithm in presence of errors will be performed.  To keep
the computation time reasonable, we chose to use the slice method
with $2\times 40$ slices per iteration for transport properties (section IV).
  Although the localization length
is not exactly the correct one, the system is still localized and enables
to study the variation of transport properties in presence of errors and 
imperfections.  For computation of the spectrum (section V), we used 
$2\times 100$ slices per iteration.

\subsection{Chebyshev polynomials}

In this approach, one uses the QFT as in the preceding methods to
shift back and forth between $\theta$ and $n$ representations.
In each representation, the relevant operator is implemented by 
using a polynomial approximation of the cosines. Since polynomials
can be implemented directly through controlled operations,
this avoids the use of additional registers.  A commonly used
polynomial approximation rests on Chebyshev polynomials.  

Chebyshev polynomials (see for example \cite{chebyshev}) 
are defined by the recurrence relation
\begin{eqnarray*}
T_0(x) & = & 1\\
T_1(x) & = & x\\
T_{n}(x) & = & 2xT_{n-1}(x) - T_{n-2}(x) \mbox{ for } n\geq2
\end{eqnarray*}
They are bounded by $-1$ and $1$ on $[-1,1]$, with their extrema smoothly
distributed over this interval. If $f(x)$ is an arbitrary function on $[-1,1]$,
and we define for $j=0\,,\dots,\,M-1$
$$c_j=\frac{2}{M}\sum_{k=0}^{M-1}f\left[\cos\left(\frac{\pi\left(k+\frac{1}{2}
\right)}{M}\right)\right]\cos\left(\frac{\pi j\left(k+\frac{1}{2}\right)}{M}
\right)$$
Then, for large $M$
$$\sum_{j=0}^{M-1}c_j T_j(x) - \frac{1}{2}c_0$$ is a very good approximation
of $f(x)$ on $[-1,1]$.\\

If we truncate this formula to order $m$:
$$\sum_{j=0}^{m}c_j T_j(x) - \frac{1}{2}c_0$$ then the error is bounded by
$\sum_{j=m+1}^{M-1}|c_k|$ and smoothly spread over $[-1,1]$.
Practically, the $c_k$'s are always rapidly decreasing, so the error term
is dominated by $|c_{m+1}|$ and we can choose a small $m$ while still
keeping a good polynomial approximation of $f(x)$.\\

Let $P(x)$ be a Chebyshev polynomial approximation of $\cos{(\pi(x+1))}$. 
If one wants to perform the operator
$ U_k=e^{-ik\cos{(p\,\hat{\theta}})} $ on a $N_H$-dimensional vector
with $N_H=2^{n_r}$ as in the preceding subsection, 
then for $p=1$
$$ U_k \approx e^{-ikP\left(\frac{\hat{\theta}}{\pi}-1\right)} $$
$U_k$ can be decomposed as a product of operators of the form $A_r(\beta)=e^{i\beta{\hat{\theta}}^r}$.

From (\ref{eq:thetadef}),
$$e^{i\beta{\theta}^r}=\prod_{j_1\dots j_r}
e^{i\beta \left(\frac{2\pi}{N_H}\right)^r d_{j_1}\dots d_{j_r} 2^{j_1+\dots +j_r}}$$
Since the $d_j$'s are binary digits, $d_{j_1}\dots d_{j_r}$ is equal to zero
unless all terms are equal to one. If we denote by $C_{j_1\dots j_r}(\phi)$
the multi-controlled phase gate, which apply the phase $\exp{(i\phi)}$
conditionally on the control qubits $j_1\ldots j_r$ (if an index is redundant, then it is counted only once),
$$A_r(\beta)=\prod_{j_1\dots j_r}
C_{j_1\dots j_r}\left(\beta \left(\frac{2\pi}{N_H}\right)^r 
 2^{j_1+\cdots +j_r}\right)$$

Since all these gates commute, and since all the gates used for the construction of $A_r$ are also present
in the development of $A_{r'}$ for ${r'}\ge r$, then all the terms of the polynomial $P$
can be applied at the same time as the term of highest order by merging similar gates.\\

\begin{figure}
\includegraphics[width=\columnwidth]{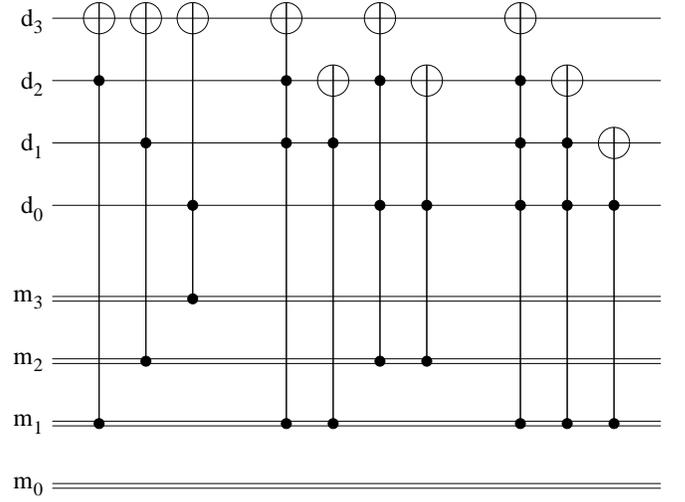}
\caption{\label{multiplication} Circuit for multiplying the quantum register
$\theta$ (simple lines) by an odd classical number $m$ (double lines)} 
\end{figure}

If $p\ne1$, then $p$ is split into $p=2^{a}m$ with $m$ odd, as in (\ref{sec:slices}).
The even part $2^a$ is dealt with by applying $U_k$ only on the $n_r-a$ first qubits.
We then multiply the register by $m$ before applying the cosine kick.
Since $m$ is co-prime with the dimension of the Hilbert space $N_H=2^{n_r}$,
this operation is unitary and can be performed without any additional qubit
(for example with the circuit in Fig.(\ref{multiplication})).\\

If we choose a Chebyshev polynomial approximation of degree $d$, then the
complexity of the algorithm is $O({n_r}^d)$.
This method is economical in qubits, and the precision of the approximation
is easy to control.  On the other hand, the complexity increases with the
precision, and this can become prohibitive for very precise simulations.
It is nevertheless quite efficient for fixed
precision computations, as can be inferred from the fact
that it is actually the method used in classical
computers to evaluate functions.

\begin{figure}
\includegraphics[width=\columnwidth]{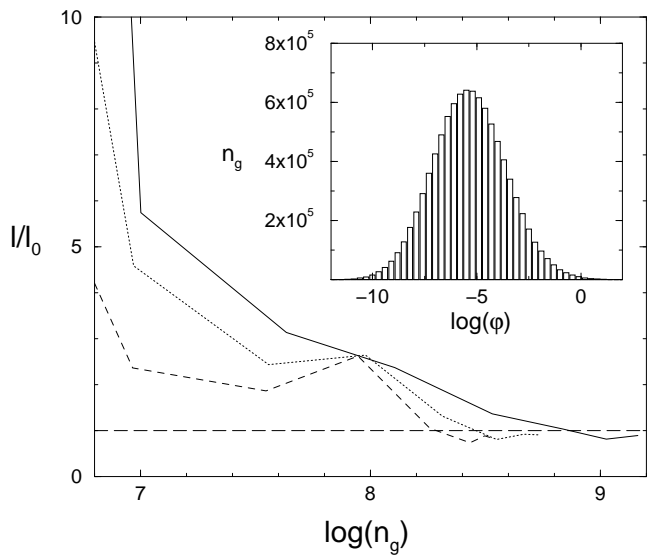}
\caption{\label{chebyshev_localisation_portes}
Localization length computed with the Chebyshev method
over exact localization length $l_0$
 as a function of number
of gates.  System parameters are the same as in 
Fig.\ref{longueur_slices}, with 
$n_r=7$ (dashed line), $n_r=8$ (dotted line), $n_r=9$ (full line)
($n_q=n_r$).
Dashed horizontal line is $l=l_0$.
Chebyshev polynomial of degree $6$ is taken, keeping gates with the
largest phases. Inset: number of gates as a function of their phase.
Logarithms are decimal.}
\end{figure}

In our numerical simulations, we found that
a Chebyshev polynomial of degree $6$ was enough to get a very
good approximation of the wave function.  This demands a much larger number
of gates than the slice method, and scales badly with $n_q$, in $n_q^6$
(here $n_q=n_r$ since there are no ancilla or workspace qubit).
However, 
some of the control-phase gates
have very small phases and are physically irrelevant. 
We can then choose a precision threshold
and simply drop all the gates with phases below this threshold. 
The distribution of the phases of the gates computing
the Chebyshev approximant of degree $6$
is displayed in the inset
of Fig.\ref{chebyshev_localisation_portes}.  

\begin{figure}
\includegraphics[width=\columnwidth]{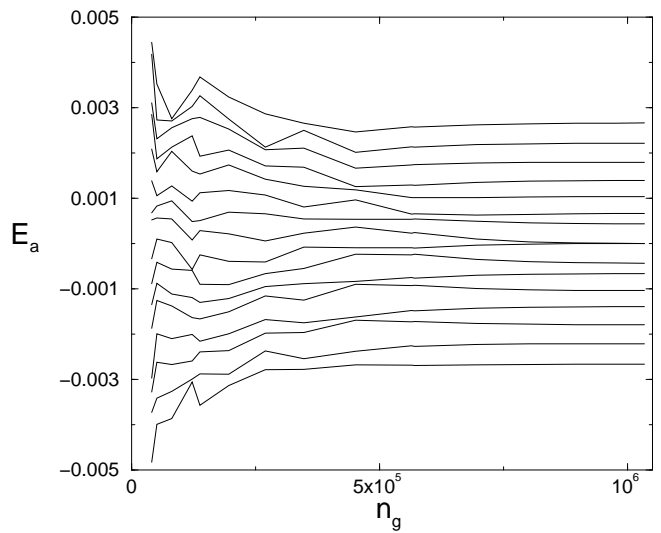}
\caption{\label{chebyshev_spectre_portes} Eigenphases of (\ref{qharper})
 as a function of number
of gates with the Chebyshev method. System is the same as in 
Fig.\ref{slices_spectre_portes}.
Chebyshev polynomial of degree $6$ is taken, keeping gates with the
largest phases. An overall phase factor (global motion of eigenvalues)
has been eliminated.}
\end{figure}

This method of approximation is investigated in 
Fig.\ref{chebyshev_localisation_portes}-\ref{chebyshev_spectre_portes}.
The localization length
as a function of number of gates is displayed
in Fig.\ref{chebyshev_localisation_portes}, 
for the same system parameters
as in Fig.\ref{longueur_slices}. 
In Fig.\ref{chebyshev_spectre_portes}, 
we display the convergence for the spectrum, in the same regime as
in Fig.\ref{slices_spectre_portes}.

In both cases, the convergence is good for maximal number of gates, showing
that the polynomial of degree $6$ is indeed a good enough approximation in
this regime of parameters.  A good accuracy is achieved
for a lower number of gates, implying that dropping the gates 
with the smallest phases can be an effective way to shorten the computation
keeping a reasonable accuracy.  Still, 
the data presented lead to the conclusion that even with the elimination of 
a large number of gates the method is clearly costlier in running time
 than the slice method 
to achieve similar precision.

\section{Transport properties: measurement and imperfection effects}

The three methods exposed above enable to simulate efficiently the
effects of the evolution operator $\hat{U}$ of the kicked Harper
model on a wave function.  This produces the wave function at a given time.
An important question concerns which quantities can be obtained
through quantum measurement of the registers, and if the whole process
including measurement is more efficient than classical computation.
A separate question but also related to practical efficiency of these
algorithms is their stability with respect to errors and imperfections
while running them on a realistic quantum computer.

In this section, we will focus on the transport properties of the
wave function. 
We recall that for the kicked Harper model, for small $K,L$ diffusion
can only takes place on the small chaotic layer of the stochastic web.
Then for larger $K,L$ there is a regime 
of parameters where all eigenstates are localized, and another regime
where localized and delocalized eigenstates coexist (see Fig.\ref{carte}).
In these different parameter regimes, 
we will show that quantities measuring localization
properties and diffusion
can be obtained on a quantum computer more efficiently than on 
a classical device, although the gain is usually polynomial.  
We will then test the resilience to errors of these quantities
obtained through the quantum algorithms, in particular through
large-scale numerical computations.
The error model
we chose corresponds to static internal imperfections. 
Indeed, physical realization of a quantum computer will never
be perfect, and some amount of disorder will always be present.
In particular, interactions between qubits, which are needed to 
build the two-qubit gates, cannot in general be totally eliminated
when they are not needed.  
These static imperfections are not linked to interaction with the outside
world; they have been shown to give important effects, which
can be larger
than the effects of noisy gates \cite{complex,qchaos,wavelet}.
To model such errors, between each gate we require that the system
evolves through the Hamiltonian

\begin{equation}
\label{hamil}
H_1 = \sum_{i} (\Delta_0+\delta_i) \sigma_{i}^z + \sum_{i} J_{i} 
\sigma_{i}^x \sigma_{i+1}^x,
\end{equation}

where the $\sigma_{i}$ are the Pauli matrices for the qubit $i$ and the second
sum runs
over nearest-neighbor qubit pairs on a circular chain.
The energy spacing between the two states of a qubit is represented 
by its average value $\Delta_0$ plus a detuning $\delta_i$
randomly and uniformly distributed in the interval 
$[-\delta /2, \delta/2 ]$.  The detuning
parameter $\delta$
gives the width of the distribution near the average value $\Delta_0$ 
and may vary from $0$ to $\Delta_0$.
The couplings $J_{i}$ represent the residual static interaction 
between qubits and is chosen 
randomly and uniformly distributed in the interval $[-J/2,J/2]$.
We make the approximation that this Hamiltonian (\ref{hamil})
 acts during a time $\tau_g$
between each gate which is taken as instantaneous.
Throughout the paper, we take in general one single rescaled
parameter $\varepsilon$
which describes the 
amplitude of these static errors, with $\varepsilon=\delta\tau_g=J\tau_g$.
To probe the transport properties of the kicked Harper model
 on a quantum computer, we chose to set $\hbar$ constant; in this way, changing
the number of qubits is equivalent to changing the size of phase space (adding
one qubit doubles the size of the phase space).  The only exception
is in the first following subsection (near-integrable regime),
where the phase space volume is constant and $\hbar$ varies with
the number of qubits.  Throughout this section, effects of
imperfections will be assessed using the slice method
to implement (\ref{qharper}).  Therefore the presence 
of one ancilla qubit implies that $n_q=n_r+1$ in all of this section.

\subsection{Near-integrable regime: stochastic web}

For $K,L$ very small, the classical system is near-integrable~:
quantum transport is dominated by the presence of invariant curves.
Motion from cell to cell can take place only by tunneling effect, or by
moving in the small chaotic zone around separatrices.  In the case $K=L$,
this small layer forms a ``stochastic web'' (see Fig.\ref{stochastic})
which extends in both $\theta$ and $n$ directions.  A wave packet started 
in this region will slowly diffuse along this web.  This process is best
seen using quantum phase space distributions, which allow direct 
comparisons between classical distributions such as the ones in
Fig.\ref{phases_harper}-\ref{stochastic} and quantum wave functions.

The Wigner function \cite{wigner,discrete} is an example of such
 quantum phase space distribution.  However, it can take negative values,
and only a smoothing over cells of
area $\hbar$ gives non negative values.  The use of a gaussian smoothing 
leads to the Husimi distribution (see e.g. \cite{husimi})
which in our case is defined by the formula:

\begin{eqnarray}
\label{husimifunction}
\lefteqn{ h(\theta,n) =}& \nonumber \\
&\sqrt{\frac{2P}{QN_{\!H}^3}} \;\;
{\displaystyle \left|\sum_{m=n-N_{\!H}/2+1}^{n+N_{\!H}/2}
\psi(m) \,e^{-\frac{\pi P}{N_{\!H} Q}(m-n)^2}
e^{2 i\pi \frac{m\Theta}{N_{\!H}}}\right|}^2 
\end{eqnarray}

where the gaussian for simplicity 
is truncated for values larger than $N_H/2$, $ \psi(m)$
is the wave function in momentum representation, $P$  (resp. $Q$) 
is the number of cells
in the momentum (resp. position) direction, $N_H=2^{n_r}$ is the dimension
of the Hilbert space,
and $\Theta=N_H\theta/(2\pi Q)$.
We note that methods to compute phase space distributions on
a quantum computer were discussed in \cite{frahm,paz,saraceno}.

\begin{figure}
\includegraphics[width=\columnwidth]{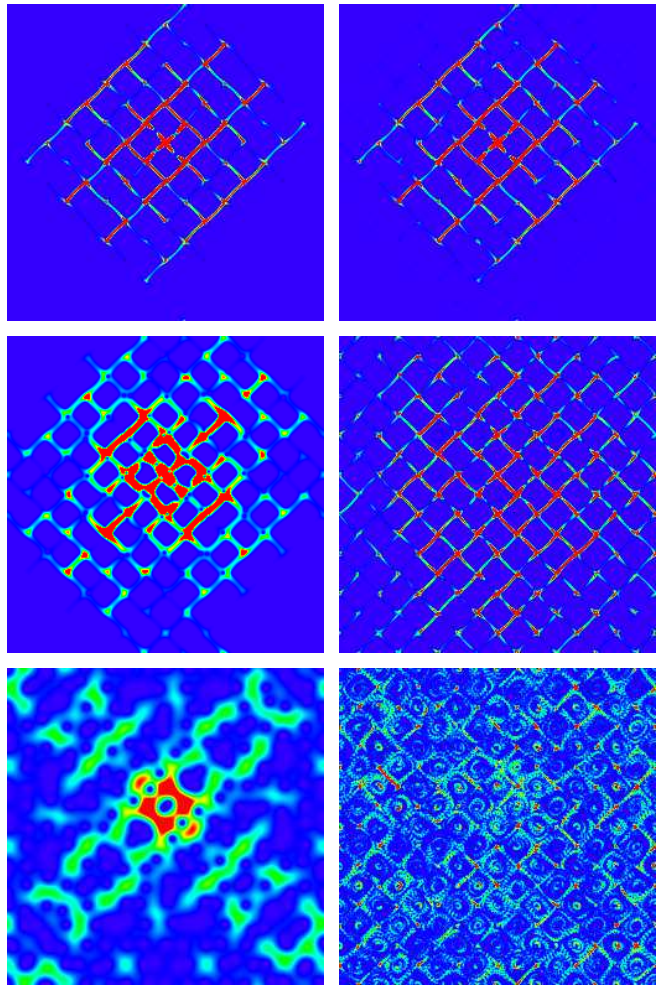}
\caption{\label{husimi} (Color online) Example of Husimi distribution of
a wave packet spreading on the stochastic web;
here $K=L=0.5$, $\hbar=2\pi\times 64/2^{n_r}$ ($8\times8$ cells),
initial state is a gaussian wave packet of area $\hbar$ 
started half a cell above the center of the figure, 
after 100 iterations using $2 \times 40$ slices per iteration.
Left: $\varepsilon=0$ and from top to bottom 
$n_r=14$, $n_r=11$, $n_r=8$ ($n_q=n_r+1$); right: $n_r=14$ and from top
to bottom $\varepsilon=10^{-6}$, $\varepsilon=10^{-5}$, $\varepsilon=10^{-4}$.
Color/grayness is related to amplitude of the Husimi function,
from zero (blue/black) to maximal value (red/white). 
Compare with the classical diffusion in Fig.\ref{stochastic}.} 
\end{figure}

In Fig.\ref{husimi} we show the spreading of a wave packet along the 
stochastic web for different numbers of qubits and different 
strengths of imperfections.   In this picture, the size of
the classical phase space is fixed, and the number of qubits
gives the value of $\hbar$.  A diffusion process is observed,
which can be related both to the classical diffusion 
on the stochastic web (Fig.\ref{stochastic}) and to the 
effect of quantum tunneling from cells to cells.  The diffusion constant
is seen from Fig.\ref{husimi} to depend on $\hbar$; it also depends on 
$K,L$ (data not shown) and is clearly different from the classical 
diffusion constant (compare the different times
in Fig.\ref{stochastic} and Fig.\ref{husimi}).  
In this near-integrable regime, the tunneling
process is quite complicated and was recently studied in \cite{ullmo}.
In the same figure, one can see that with moderate levels of 
imperfections the exact Husimi distribution is well 
reproduced by the algorithm.

To probe transport properties in this regime, one can start a wave packet
in the stochastic web and let it evolve.  After a certain number of 
time steps, the diffusion constant can be obtained from measurement 
of the wave function.  As the number of components of the 
wave function or of the Husimi distribution becomes exponentially large
as $n_r$ increases, the best way is to use coarse-grained measurements~:
measuring only the first qubits adds up the amplitudes squared
 of many neighboring
components and limits the number of measurements to a fixed value.
This can be done to the wave function directly in the momentum or position
representation, or to the Husimi function provided all the values are kept
on a quantum register.  For example, the Husimi-like function developed
in \cite{frahm} can be obtained by modified Fourier transform from the
wave-function, and allows the use of coarse-grained measurements.
If one starts a wave packet on the stochastic web, it will diffuse
according to the law $\langle s(t)^2\rangle \approx D_s t$, with $s$ being a 
distance
in phase space and $D_s$ the diffusion constant.  Performing time evolution
up to a time $t^*$ requires $t^*$ quantum operations multiplied by
logarithmic factors.  At this stage, a fixed number of coarse-grained 
measurements is enough to give an approximation of $D_s$.
On a classical computer, one can truncate 
the Hilbert space up to the maximal dimension effectively used in the
calculation, which is of the order $\sqrt{t^*}$. Propagating the
wave packet will cost $t^* \sqrt{t^*}$ classical operations, after which 
$D_s$ can be obtained.  Therefore the quantum computation is 
polynomially faster than the classical one.  Methods which use
an ancilla qubit to measure the value of phase space distributions at
a given point such as the ones in \cite{paz,saraceno} will necessitate 
extra measurements since they cannot be used to perform coarse-grained 
measurements efficiently.  Still, by reducing $K,L$ as $n_r$ is
increased, one can keep the number of large components of the Husimi
function of the wave packet of order $N_H$ (instead of $N_H^2$).  In this case, 
the Husimi function measured on the ancilla qubit of \cite{saraceno} 
is efficiently measurable.  This is formally an exponential gain
over direct classical simulation since measuring one component
of the Husimi distribution at a fixed time $t$ will be logarithmic
in $N_H$.  The same happens for coarse-grained measurements at fixed
$t$.  Still, as $\hbar$ goes to exponentially small values the
dynamics for fixed $t$ will become very close to the classical one,
so it is unclear which new information can be gained this way.

\begin{figure}
\includegraphics[width=\columnwidth]{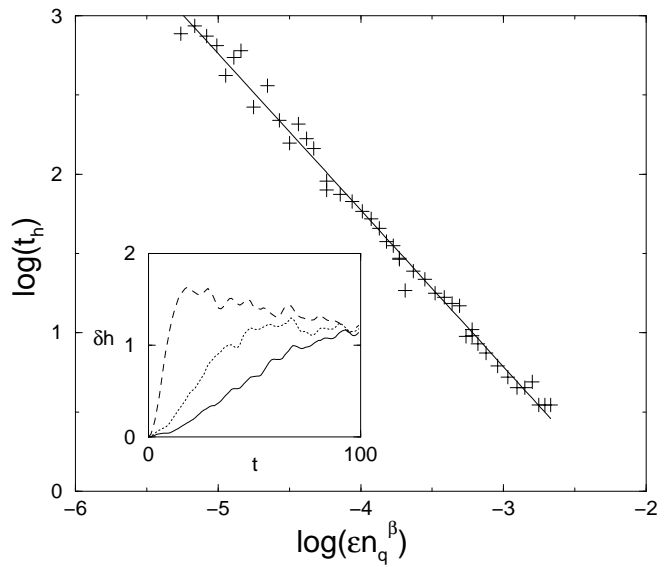}
\caption{\label{errorhusimi}Effects of imperfections on the
 Husimi distribution of
a wave packet spreading on the stochastic web~; 
here $K=L=0.5$, $\hbar=2\pi\times 64/2^{n_r}$ ($8\times8$ cells),
initial state is a gaussian wave packet of area $\hbar$ 
started half a cell above the center of the Fig.\ref{stochastic}, 
iterations are made by the slice method
using $2 \times 40$ slices per iteration.  Straight line is the
law (\ref{thusimi}) with $\alpha=1$ and $\beta=1.23$.
Crosses corresponds to various values of $\varepsilon$ 
($10^{-6}\leq \varepsilon \leq 10^{-4}$)
and $n_r$ ($5\leq n_r \leq 14$, with $n_q=n_r+1$), averages were made
over all Husimi components inside the stochastic web and
up to $100$ realizations of disorder for each $\varepsilon$ value.
Inset~: average relative error of the Husimi function 
$\delta h= \langle |h_{\varepsilon} - h_0|\rangle /\langle h_0\rangle$
 on the stochastic web for $\varepsilon=10^{-4}$ (dashed line),
$\varepsilon=10^{-4.5}$ (dotted line), $\varepsilon=10^{-5}$ (full line), 
$n_r=10$
($n_q=n_r+1$).
Average is taken over all Husimi components inside the stochastic web and
$10$ realizations of disorder.  Logarithms are decimal.
} 
\end{figure}

To clarify the stability of these algorithms with respect to errors,
in Fig.\ref{errorhusimi} we show quantitatively the
effects of imperfections on the Husimi distributions for a
wave packet spreading on the stochastic web
for various numbers of qubits and imperfection strengths.  We computed 
the time scale $t_h$ for various parameter values, 
$t_h$ being the time (number of iterations) for which
the error on the Husimi functions is half the mean value of that function
on the stochastic web.

The numerical data
suggest the law 
\begin{equation}
\label{thusimi}
t_h\approx C_h/(\varepsilon^{\alpha} n_q^{\beta}) 
\end{equation}
with $\alpha =1.02 \pm 0.02$ (compatible with 
$\alpha=1$) and $\beta=1.23\pm 0.09$ with $C_h\approx 0.007$.
  This law is polynomial in both $\varepsilon$
and $n_q$, which indicates that even though
individual values of the Husimi function can be exponentially
small, the effect of imperfections remains small compared to these
individual values for a polynomial time.
This means that such quantities can be reliably obtained in presence
of moderate levels of imperfections. More work is needed
to understand the precise origin of the law (\ref{thusimi}).
We note that in \cite{Levi} where 
random noise in the quantum gates were used as main source of errors
a similarly polynomial (but different)
law was found for the relative error on the Husimi function.

\subsection{Localized regime}

\begin{figure}
\includegraphics[width=\columnwidth]{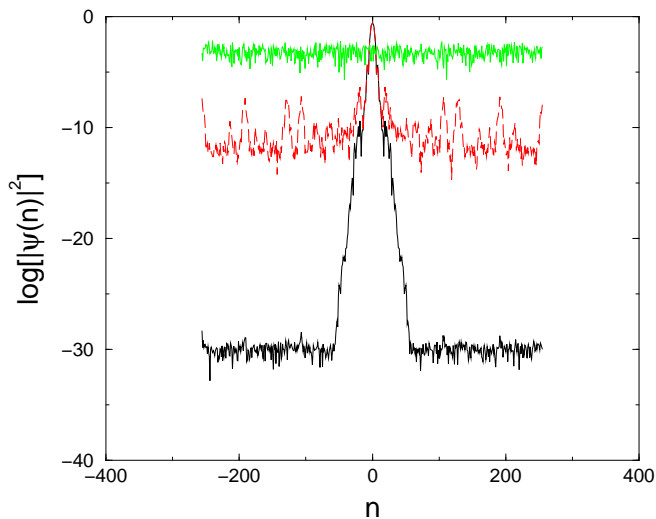}
\caption{\label{psiK1} (Color online) 
Example of wave function in the localized regime.
here $K=1$, $L=5$, $\hbar/(2\pi)=(13-\sqrt{5})/82$ (actual value is the 
nearest fraction with
denominator $2^{n_r}$),
initial state is $|\psi_0\rangle=|0\rangle$,
after 1000 iterations using $2 \times 40$ slices per iteration,
$n_r=8$ ($n_q=n_r+1$), from bottom to top
$\varepsilon=0$ (black, solid line), $\varepsilon=10^{-7}$
(red, dashed line), $\varepsilon=10^{-3}$ (green, solid line).
In the center, the first two curves are superposed and indistinguishable.
Logarithm is decimal.} 
\end{figure}

When $K$ is large enough for the chaotic zone to
take most of the classical phase space, a classical
particle will propagate diffusely in phase space.  
In contrast, for moderate
values of the parameter $K$, all the eigenstates of 
the evolution operator $\hat{U}$ of (\ref{qharper}) are localized
(see Fig.\ref{carte}).  This localization is a purely
quantum phenomenon due to interference effects and similar
to the Anderson localization of electrons in solids.
In this parameter regime, an initial wave packet will have projections
on only a small number of exponentially localized eigenstates.
Thus after a few iterations of the map, the wave packet will stop
spreading and stay in a region of momentum space
of size given by the localization length.  An example of such
a wave function is shown in Fig.\ref{psiK1}. 

In this regime, it is possible to measure the localization length $l$
efficiently.
Indeed, most of the probability is
concentrated in a domain of size $l$.  If one performs a 
coarse grained measurement of the wave function, i.e. only the most significant
qubits are measured, the number of measurements will set the 
precision in units of $l$. Thus once the desired relative precision is fixed,
the number of measurements is independent of $l$ or $n_q$.  Nevertheless,
if one starts from an easily prepared initial wave packet, for example 
on a single momentum state, one has to evolve it long enough to
reach a saturation regime where the wave function is spread on a domain of size
$\approx l$.  Classically, in the parameter regime where the system
is chaotic, the dynamics is diffusive $\langle n(t)^2\rangle \approx Dt$
with a diffusion constant $D$ which depends on parameters.  One can expect the
wave packet to follow for short time this diffusive behavior which will
stop when a spreading comparable to the
localization length is reached.  In this case, the wave packet needs to be
evolved until a time $t^{*}\approx l^2/D$.  Classically, one needs 
to evolve a vector of dimension $\sim l$
until the time $t^{*}$; this needs 
$\sim l^3$ classical operations.  On a quantum computer,
once the precision is set the three algorithms above need only a logarithmic 
number of gates to perform one iteration, so the total number of gates
is $\sim l^2$.  This gives a polynomial improvement for the quantum algorithm.
It is known that in the delocalized phase, the wave packet can spread 
ballistically for some regimes of parameter.  If this extends to short times
and to the localized regime, 
then the gain becomes quadratic.  

In Fig.\ref{psiK1}, an example of a localized wave function is shown 
for different imperfection strengths.  At $\varepsilon=0$, the exponential 
localization is clearly visible, the exponential decay being leveled off
at very small values ($\approx 10^{-30}$) only by numerical roundoff.  For
larger values of $\varepsilon$, the localized peak is still visible with the 
correct amplitude, but a larger and larger background is visible, 
until the peak disappears.

To analyze in a more precise way
the effects of imperfections, 
we have to specify the observable
that is used to get the localization length.
On a classical computer, different data analysis can be used to
calculate the localization length from knowledge of the wave function.
A first way consists in extracting the second moment of the wave function
$\langle (\Delta n)^2 \rangle$, which gives an estimate of $l$
once the saturation regime is reached.  One can also compute the 
inverse participation ratio (IPR)
$\xi=1/\Sigma_n |\psi(n)|^4$.  For a wave function 
uniformly spread over $M$ states this quantity is equal to $M$, and therefore
it also gives an estimate of the localization length.  At last, $l$ can
be measured directly
by fitting an exponential function around maximal values of $\psi$.

\begin{figure}
\includegraphics[width=\columnwidth]{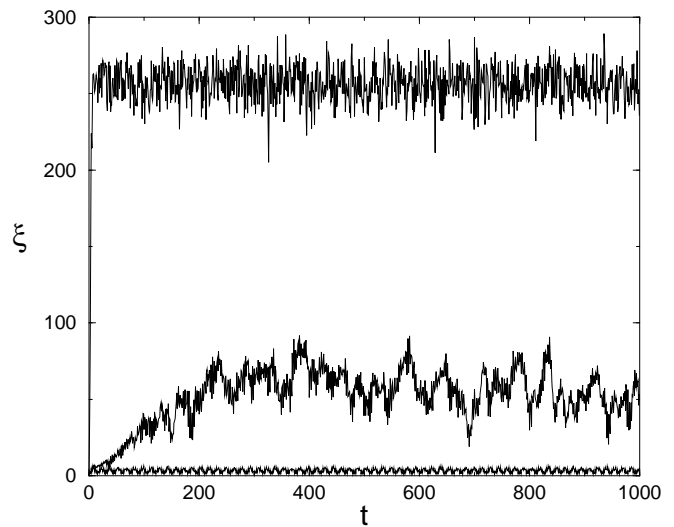}
\caption{\label{ipr_t_localise} Example of
IPR with imperfections as a function of time,
in the localized regime.  Parameters values are the same as in Fig.\ref{psiK1},
$n_r=8$ ($n_q=n_r+1$), from bottom to top
$\varepsilon=0$ , $\varepsilon=10^{-7}$,
 $\varepsilon=10^{-4}$,
$\varepsilon=10^{-3}$. Data from $\varepsilon=0$ and 
$\varepsilon=10^{-7}$ are indistinguishable.} 
\end{figure}

For an exact wave function, all three quantities give similar results. 
On a quantum computer, they may have very different behavior with respect to 
imperfection strength.
Indeed, it was shown in general \cite{Levi,loclength} that the 
second moment is exponentially sensitive to the number of qubits in presence
of imperfections, making it a poor way to get information about transport
properties.  The IPR was shown \cite{loclength} to be polynomially sensitive
to both number of qubits and imperfection strength.  Still, the IPR may be 
difficult to measure directly on a quantum computer.
On the other hand, the direct measurement
of $l$ by fitting an exponential curve on a coarse-grained measure of
the wave function was shown in \cite{loclength} to be an effective way
to extract $l$ from a quantum computation of the wave function.  
It is therefore interesting to study the behavior of both latter quantities
with respect to imperfections.
 
In Fig.\ref{ipr_t_localise}, the time evolution of the IPR is shown
for different values of the imperfection strength.  For $\varepsilon=0$,
the wave packet first spreads for $t<t^{*}$ then the IPR becomes approximately
constant and close to the localization length.  For larger values of 
$\varepsilon$, 
the wave packet spreads to much larger parts of phase space, 
but the IPR still saturates after some time to a value which depends on
$\varepsilon$ and $n_q$. 

\begin{figure}
\includegraphics[width=\columnwidth]{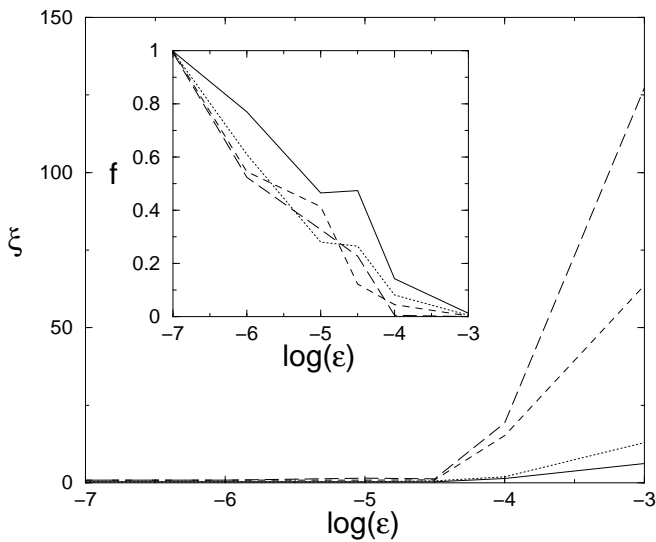}
\caption{\label{localisation_ipr_erreur_inset} 
IPR as a function of imperfection
strength in the localized
regime.  Parameters values are the same as in Fig.\ref{psiK1},
 with $n_r=7$ (full line),
$n_r=8$
(dotted line), $n_r=9$ (dashed line),
$n_r=10$ (long dashed line) ($n_q=n_r+1$). 
Averages were made over up to $10$ realizations of disorder.
Inset: Fidelity as a function of imperfection
strength in the localized
regime, with same parameter values and line codes as in the main figure. 
Logarithms are decimal.} 
\end{figure}

\begin{figure}
\includegraphics[width=\columnwidth]{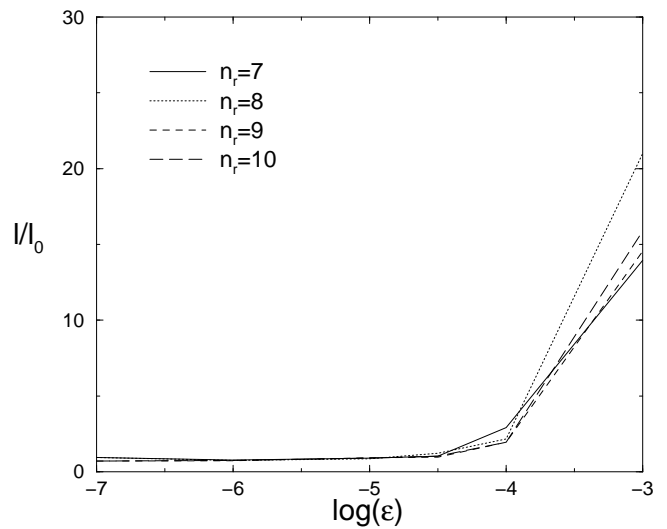}
\caption{\label{slices_localisation_longueur_erreur} 
Localization length as a function of imperfection
strength in the localized
regime.  Parameters values are the same as in Fig.\ref{psiK1}, with
averages made over up to $10$ realizations of disorder.
Logarithm is decimal.}
\end{figure}

The average value of this saturation value
is shown in 
Fig.\ref{localisation_ipr_erreur_inset} as a function of $\varepsilon$ for
different values of $n_q$.
Fig.\ref{slices_localisation_longueur_erreur} shows the localization length
obtained from curve-fitting for the same wave functions.
For large enough values of $\varepsilon$, 
the IPR grows very quickly, in a manner which seems exponentially
dependent on $n_q$. The result of
the curve-fitting strategy is roughly similar, but shows an intermediate regime
($\varepsilon \approx 10^{-4}$ for our data) where it is still reasonably close
to the exact value while the IPR is already quite far off.
 This can be understood qualitatively
from the data shown on Fig.\ref{psiK1}.
Indeed, the effect of moderate static imperfections is to create a larger
and larger background over which the localization peak is superimposed.
The IPR is sensitive to the presence of this background, while by
its very definition the curve-fitting strategy isolates the localization peak
from the background and is therefore more robust.
  The data presented in Fig.\ref{slices_localisation_longueur_erreur}
show that
this peak keeps its shape with relatively good accuracy until its final 
disappearance, even though a large chunk of its amplitude has been transfered
elsewhere by imperfections.  
The inset of Fig.\ref{localisation_ipr_erreur_inset} shows the fidelity
of the same wave functions.
($f(t)=|\langle\psi(t)|\psi_{\varepsilon}(t)\rangle|^2$ where 
$|\psi(t)\rangle$ is the
exact wave function and $|\psi_{\varepsilon}(t)\rangle$ the one in presence of
imperfections).  It is interesting to note that the localization length and IPR
can be quite well reproduced even for values of $\varepsilon$ where 
the fidelity is already quite low.

A more precise analysis can be 
developed from the effect of imperfections on the
eigenstates of the unperturbed evolution operator
$\hat{U}$ in (\ref{qharper}).
  These eigenstates $|\psi_a\rangle$ can be written as
a sum over momentum states $|m\rangle$, 
which coincide with quantum register states
of the quantum computer when the system is in momentum representation~:

\begin{equation}
\label{eigenstate}
|\psi_a\rangle = \sum_{m=1}^{N_H} c_a^m |m\rangle
\end{equation}

In the localized regime, the eigenstates $|\psi_a\rangle$ are localized
with localization length $l$,
therefore the $c_a^m$ are significant only for $\sim l$ values of $m$,
with average value $1/\sqrt{l}$.  Using perturbation theory, one
can estimate the typical matrix element of
the imperfection Hamiltonian (\ref{hamil}) between eigenstates.
For the first term of (\ref{hamil}), this gives:

\begin{eqnarray}
\label{perturb}
V_{typ}&\sim& \left|\langle \psi_b | \sum_{i=1}^{n_q} 
\delta_i \widehat{\sigma_{i}^z}\tau_g n_g
  |\psi_a\rangle \right|          \nonumber \\
&\sim & \tau_g n_g \left|\sum_{m,n=1}^{N_H} c_a^m {c_b^n}^{\ast} \langle n |
\sum_{i=1}^{n_q}\delta_i \widehat{\sigma_{i}^z}  |m\rangle \right|
\end{eqnarray}

where $N_H=2^{n_r}$ is the dimension of Hilbert space on which $\hat{U}$ acts,
$\tau_g$ is the time for one gate, and the term due to $\Delta_0$ 
in (\ref{hamil}) is not taken into account since it can be eliminated easily.
This estimate (\ref{perturb})
 is an approximation, since the action of (\ref{hamil})
is separated from the action of $\hat{U}$ and in reality they are intertwined
and do not commute.
In (\ref{perturb}), only $\sim l$ neighboring quantum register states
are coupled 
through $n_q$ terms of different detuning $\delta_i$ (with random sign).
This term therefore gives on average 
$\varepsilon n_g \sqrt{n_q} /\sqrt{l}$.  
The second term of (\ref{hamil}) in the same approximation
will be the sum of 
$n_q$ terms, each coupling one state $|m\rangle$
with another state differing by two
neighboring qubits $|n\rangle=|m+r\rangle$.  So a state $|\psi_a\rangle$
is coupled significantly only to states $|\psi_b\rangle$ localized
at a distance $r$  in momentum from $|\psi_a\rangle$.  Therefore the same
estimate applies, and overall one can estimate 
$V_{typ} \sim \varepsilon n_g \sqrt{n_q}/\sqrt{l}$.

\begin{figure}
\includegraphics[width=\columnwidth]{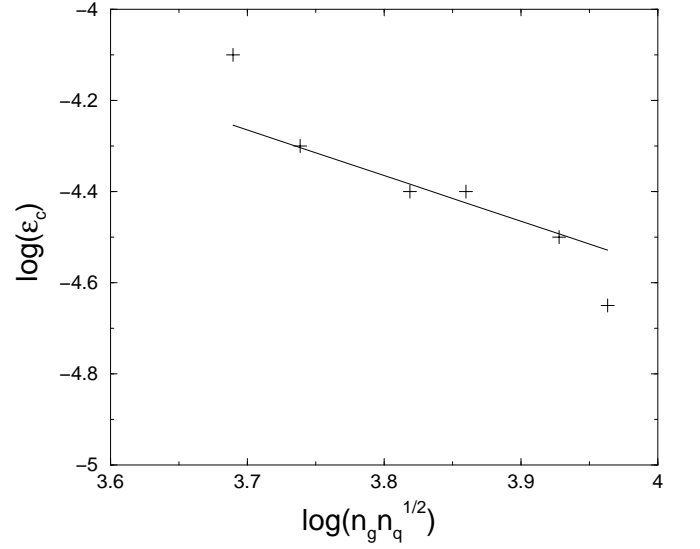}
\caption{\label{epsilonc_K2} Critical
value of $\varepsilon$ (error strength) as a function of parameters
for $K=2,L=27$, with
other parameter values the same as in Fig.\ref{psiK1}.
 $\varepsilon_c$ is defined by a saturation value
of IPR twice the unperturbed value.  Averages were made with up
to $10$ realizations of disorder.
Solid line is the formula (\ref{thresholdK2}). Logarithms are decimal.} 
\end{figure}

One can suppose that the IPR will become large when
perturbation theory breaks down.  This happens when $V_{typ}$
is comparable to the distance between directly coupled states $\Delta_c$.
From the arguments above, one expects that one state is coupled
to $\sim l$ states so that this distance is $\Delta_c\sim 1/l$.  The threshold
when IPR or localization length become large is therefore
$V_{typ}\sim \Delta_c$
which corresponds to:

\begin{equation}
\label{thresholdK2}
\varepsilon_c \approx C_1/(n_g \sqrt{n_q}\sqrt{l})
\end{equation}
where $C_1$ is a numerical constant and $n_g$ is number of gates per iteration,
$n_q$ number of qubits, $l$ the localization length.  Fig.\ref{epsilonc_K2}
is compatible with this scaling, with $C_1/\sqrt{l}\approx 0.3$.  
We note that this threshold
is similar to the threshold for the transition to quantum chaos
presented for a quantum computer not running an algorithm in \cite{qchaos}.

\begin{figure}
\includegraphics[width=\columnwidth]{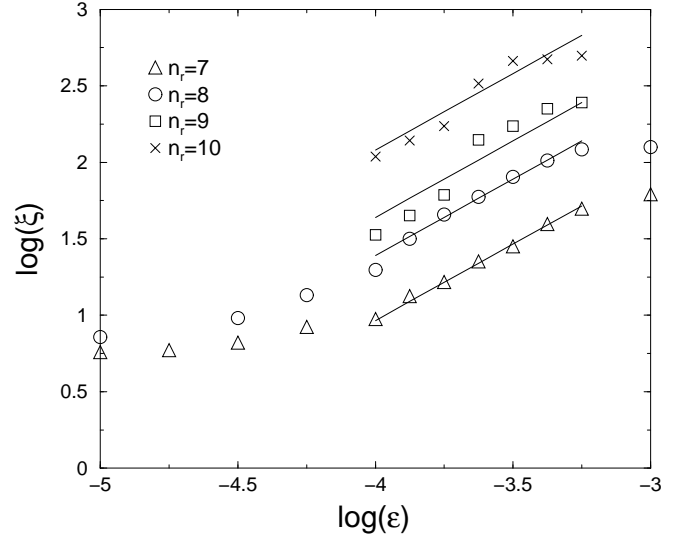}
\caption{\label{IPR_K2} IPR as a function of 
$\varepsilon$.
Parameters values are the same as in Fig.\ref{epsilonc_K2}.
Solid lines correspond to the dependence $\xi \propto \varepsilon$ . 
Logarithms are decimal.} 
\end{figure}

When perturbation theory breaks down, it is usually expected from
earlier works on quantum many-body physics \cite{qchaos,manybody}
that the system enters
a Breit-Wigner regime where the local density of states is a Lorentzian
of half-width $\Gamma\approx 2\pi |V_{typ}|^2/\Delta_c$ according to the Fermi
golden rule.  This implies that the IPR grows like 
$\Gamma/\Delta_n \sim\varepsilon^2 n_g^2 n_q N$, 
where $\Delta_n\sim 1/N_H \sim 1/N$ 
is the mean level spacing ($N=2N_H$ since there is an ancilla qubit). This is
not confirmed by the data shown on Fig.\ref{IPR_K2}, which suggest
that the IPR scales like $\varepsilon$.  This indicates that in our system
we are in a regime different from the golden-rule (Breit-Wigner) regime.

Such a regime is present for large perturbation strength in
many-body systems.  It is indeed known that for large enough values
of the couplings, the system leaves the golden rule regime
and enters a new regime where the local density of states
is a gaussian of width given by the variance $\sigma$.  The
variance can be approximated by 
$\sigma^2\sim \sum_{b\neq a} V_{typ}^2 \sim \varepsilon^2 n_g^2 n_q$.
In this regime the IPR is given by 
$\sigma/\Delta_n \sim \varepsilon n_g \sqrt{n_q} N$,
which is consistent with data from 
Fig.\ref{localisation_ipr_erreur_inset} and
Fig.\ref{IPR_K2}.  This regime
is known to supersede the golden rule regime for $\Gamma>\sigma$, 
which should therefore be the case for our system.
This implies that the relevant time scale for the system
to remain close to the exact one is $1/\sigma$.  For the largest
values of $n_q$, the data in Fig.\ref{IPR_K2} show some departure
from this law, which may be due simply to statistical
 fluctuations (the averaging
is made over more instances for smaller $n_q$), or a shift toward
the golden rule regime for large $n_q$.

The scaling laws obtained in this regime show that for $\varepsilon < \varepsilon_c$,
with $\varepsilon_c$ given by (\ref{thresholdK2}),
the system is still localized in presence of imperfections, and
the localization length is close to the exact one.
In this case, the localization length is correct for very long times,
much longer than for example the fidelity decay time.
For larger $\varepsilon$, the system with imperfections is delocalized.
We still expect it to be close to the exact one up to a time 
$\sim 1/\sigma\sim1/(\varepsilon n_g \sqrt{n_q})$.

\subsection{Partially delocalized regime}

\begin{figure}
\includegraphics[width=\columnwidth]{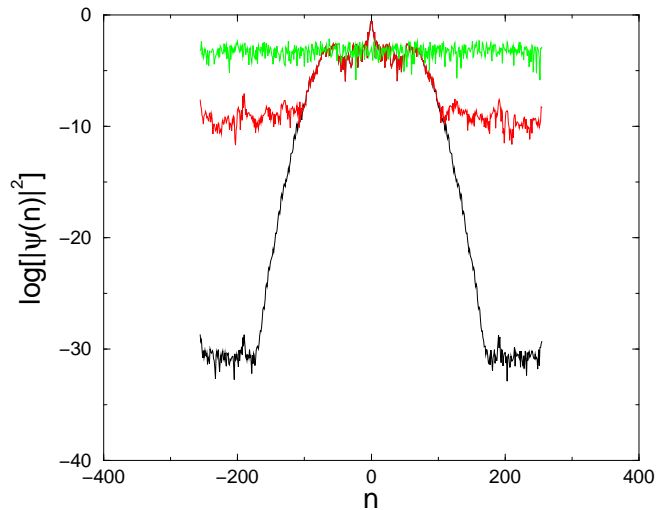}
\caption{\label{psiK2} (Color online)
Example of wave function in the partially delocalized
 regime.
Here $K=2$, $L=5$,  $\hbar/(2\pi)=(13-\sqrt{5})/82$ (actual value is the 
nearest fraction with
denominator $2^{n_r}$),
initial state is $|\psi_0\rangle=|0\rangle$,
after 100 iterations using $2\times 40$ slices per iteration,
$n_r=8$ ($n_q=n_r+1$), from bottom to top
$\varepsilon=0$ (black, solid line), $\varepsilon=10^{-7}$
(red, dashed line), $\varepsilon=10^{-3}$ (green, solid line).
In the center, the first two curves are superposed and indistinguishable.
Logarithm is decimal.} 
\end{figure}

For larger values of $K$ at $L$ fixed, the system enters a partially 
delocalized region.
In this regime, there is a coexistence of localized and delocalized 
eigenstates.  An initial wave packet will have significant projections on 
all delocalized eigenstates but only on neighboring localized eigenstates.
After a certain number of time steps (kicks) the part corresponding to 
delocalized states will spread in all the system, while the localized part
will remain close to the initial position.  Fig.\ref{psiK2} shows an example
of a wave packet initially at $n=0$ after $100$ iterations in this regime,
displaying
 an exponential peak corresponding to localization
 superimposed on a plateau which spreads with time to larger
and larger momentum.  It is known that the spreading of the wave packet 
in this regime (for large enough time) is ballistic away from the line $K=L$
and diffusive on this line.

In this regime, as above a coarse grained measurement
can give the localized part with moderate accuracy, thus enabling to
compute the localization length.  As in the preceding part, the gain over
classical computation will be polynomial. As concerns the delocalized part of 
the wave function, 
it seems at first sight that getting
information on it is difficult, since it takes
very long time to reach its saturation distribution (it has to spread
diffusively or ballistically
through the whole system), and this distribution itself
is spread over the exponentially large system.  Still, after a time large
enough for the wave packet to spread beyond the localization length,
the structure of the wave function can be seen very clearly from coarse-grained
measurements whose number is on the order of the localization length.
Once such coarse grained measurement has been performed, and
the localization length 
found by fitting an exponential function around the maximum, 
the relative importance of the plateau can be found by subtracting the
localized part.  Even though the plateau has not yet reached its final 
distribution, its integrated probability is related to
 the number of eigenstates
which are delocalized at these parameter values.  This information enables
to monitor the transition precisely for different values of $K$ and $L$, 
a non trivial information as seen from Fig.\ref{carte}.
  The number of operations
for classical and quantum algorithms are the same as for the localization 
length, and therefore the same polynomial gain can be expected.
Another quantity which can be readily obtained is the quantum diffusion 
constant. Indeed, away from the line $K=L$, it is known that a quantum wave
packet initially localized in momentum will spread anomalously (ballistically)
with the law $\langle n^2(t)\rangle \approx D_a t^2$.  Classically, 
estimating the diffusion constant requires to simulate the system until 
some time $t^*$.  This requires to evolve $\sim t_*$ quantum states until
the time $t^*$, making the total number of operation $\sim (t^*)^2$.  On a
quantum computer, one time step requires a logarithmic number of operations,
so the total number of operations is $\sim t^*$ ($t^*$ iterations 
followed by a constant number of coarse-grained measurements),
a quadratic gain compared to 
the classical algorithm.  Close to the line $K=L$, the quantum diffusion 
becomes normal with the law $\langle n^2(t)\rangle \approx D_n t$.  
In this regime, the same computation gives a number of operation 
$\sim (t^*)^{3/2}$ classically compared to $\sim t^*$ 
for the quantum algorithm,
with still a polynomial gain.  Such computations can give quite interesting
results,
in particular to specify precisely which kind of diffusion is present 
in the vicinity of the line $K=L$, a question which is not definitively 
settled.

\begin{figure}
\includegraphics[width=\columnwidth]{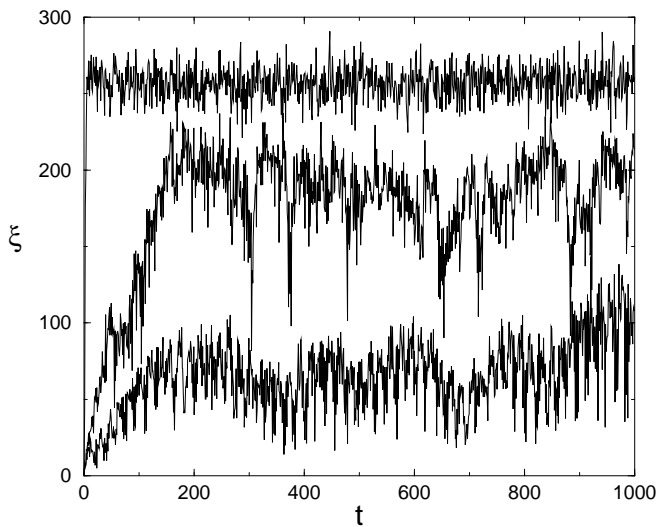}
\caption{\label{ipr_t_delocalise_K4L5} Example of IPR in presence of
imperfections as a function of time in the transition regime.
Here $K=4$, $L=5$, $\hbar/(2\pi)=(13-\sqrt{5})/82$ (actual value is the 
nearest fraction with
denominator $2^{n_r}$),
initial state is $|\psi_0\rangle=|0\rangle$
with $2\times 40$ slices per iteration,
$n_r=8$ ($n_q=n_r+1$), from bottom to top
$\varepsilon=0$, $\varepsilon=10^{-7}$,
$\varepsilon=10^{-4}$, $\varepsilon=10^{-3}$.
 Data from $\varepsilon=0$ and 
$\varepsilon=10^{-7}$ are indistinguishable.
} 
\end{figure}

Effects of different strengths of imperfections can be seen in Fig.\ref{psiK2}.
For moderate values of $\varepsilon$, a flat background of larger and larger
amplitude is created by the imperfections.  When this background reaches
the values of the plateau due to delocalized states, information on
these delocalized states is lost, but the localized peak remains until 
$\varepsilon$ is large enough to destroy it.  This is visible also on 
Fig.\ref{ipr_t_delocalise_K4L5} which displays the time evolution
of a wave function in this transition region.  The data for $\varepsilon=0$  
show the spreading of the wave packet due to delocalized eigenstates; the IPR
does not reach the dimension of Hilbert space since part of the amplitude 
does not spread due to localized states.  For intermediate values of 
$\varepsilon$, the spreading concerns more and more of the total amplitude,
increasing the IPR, until a large enough value of $\varepsilon$ is reached and 
the wave function is completely delocalized.

In this regime, the analysis of the preceding section should be modified.
Indeed, a certain fraction $\beta$ 
of the Floquet eigenstates $|\psi_a\rangle$ of 
$\hat{U}$ (unperturbed) in (\ref{qharper}) are not localized.
For these delocalized states, the $c_a^m$ of (\ref{eigenstate})
have small nonzero values $\sim 1/\sqrt{N_H}$ for all $m$.
The estimation 
$V_{typ} \sim \varepsilon n_g \sqrt{n_q}/\sqrt{l}$
for the typical matrix element of
the imperfection Hamiltonian (\ref{hamil}) between eigenstates 
$|\psi_a\rangle$ and $|\psi_b\rangle$ remains correct only if
$|\psi_a\rangle$ and $|\psi_b\rangle$ are both localized.

If $|\psi_a\rangle$ and $|\psi_b\rangle$ are both delocalized, 
one has $c_a^m \sim c_b^n \sim 1/\sqrt{N_H}$ in (\ref{perturb}) for
most $m,n$.  This implies that the quantities $\sum_{m=1}^{N_H} c_a^m c_b^m$,
previously of order $1/\sqrt{l}$ becomes $\sim 1/\sqrt{N_H}$ (sum of $N_H$ 
terms of order $\sim 1/N_H$ with random signs).  This modifies the estimate
for $V_{typ}$: with the same reasoning as in the localized case,
one has $V_{typ}\sim \varepsilon n_g \sqrt{n_q}/\sqrt{N_H}$.

If one of the states $|\psi_a\rangle$ and $|\psi_b\rangle$ is localized
and the other one delocalized, then $\sum_{m=1}^{N_H} c_a^m c_b^m$
is the sum of $l$ terms of order $\sim 1/(\sqrt{l} \sqrt{N_H})$ with random
signs, which is of order $\sim 1/\sqrt{N_H}$. 
This gives the same estimate $V_{typ}\sim \varepsilon n_g \sqrt{n_q}/\sqrt{N_H}$
for the matrix element as if both states were delocalized.

\begin{figure}
\includegraphics[width=\columnwidth]{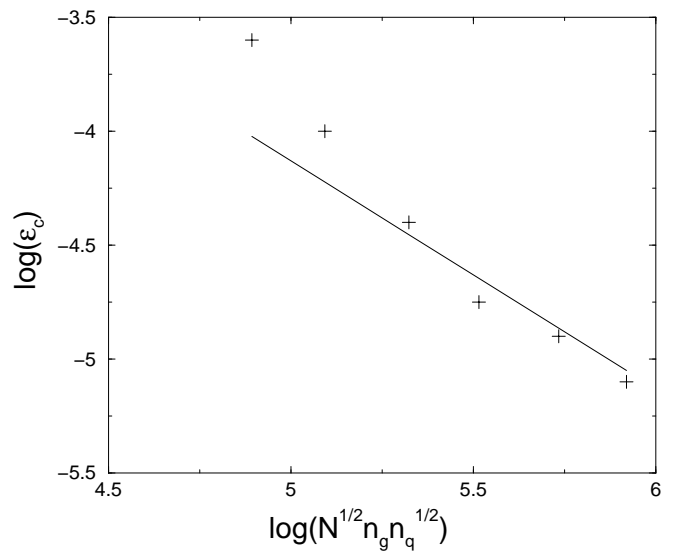}
\caption{\label{epsilonc_K10} Critical
value of $\varepsilon$ (error strength) as a function of parameters
for $K=10$, $L=27$ with
other parameter values the same as in Fig.\ref{psiK2}.
 $\varepsilon_c$ is defined by a saturation value
of IPR twice the unperturbed value.  Averages were made with up
to $10$ realizations of disorder.
Solid line is the formula (\ref{thresholdK10}). Logarithms are decimal.} 
\end{figure}

Therefore if a proportion $\beta$ of the unperturbed Floquet
eigenstates are delocalized, both localized and delocalized eigenstates will
have matrix elements of order $V_{typ}\sim \varepsilon n_g \sqrt{n_q}/\sqrt{N_H}$
with $\beta N_H$ other eigenstates.  This will be the dominant effect, since
these couplings lead perturbation theory to break down much earlier
than for the purely localized system.  Indeed,  $V_{typ}$
is comparable to the distance between directly coupled states 
$\Delta_c \sim 1/N_H\sim 1/N$ (since $N=2N_H$) for
$\varepsilon n_g \sqrt{n_q}/\sqrt{N} \sim 1/N$, which corresponds
to: 

\begin{equation}
\label{thresholdK10}
\varepsilon_c \approx C_2/(n_g \sqrt{n_q}\sqrt{N})
\end{equation}
where $C_2$ is a numerical constant, $n_g$ is the 
number of gates per iteration,
$n_q$ the 
number of qubits and $N=2^{n_q}$ the dimension of the Hilbert space
of the quantum computer.  
Fig.\ref{epsilonc_K10}
is compatible with this scaling, with $C_2\approx 7.4$. 

In this regime, the
critical interaction strength drops therefore {\em exponentially}
with the number of qubits, in sharp contrast with the localized regime.
This effect has been noted for a different system in \cite{hyper},
and is similar to the enhancement of weak interaction in heavy nuclei
\cite{SF}.  The physical mechanism is that the much smaller coupling term
between states is compensated by the even smaller distance in energy 
between coupled states.
This result implies that even for moderate number of qubits, 
a small interaction strength is enough to modify enormously
the long-time behavior of the system: saturation values of the IPR 
are very much affected by the perturbation, much more so than in the
localized regime.  However, for short time the behavior of the
system should be close to the unperturbed one, implying that 
the measures suggested to get interesting information, such as
relative size of the plateau and diffusion constants can still be accessible.

\begin{figure}
\includegraphics[width=\columnwidth]{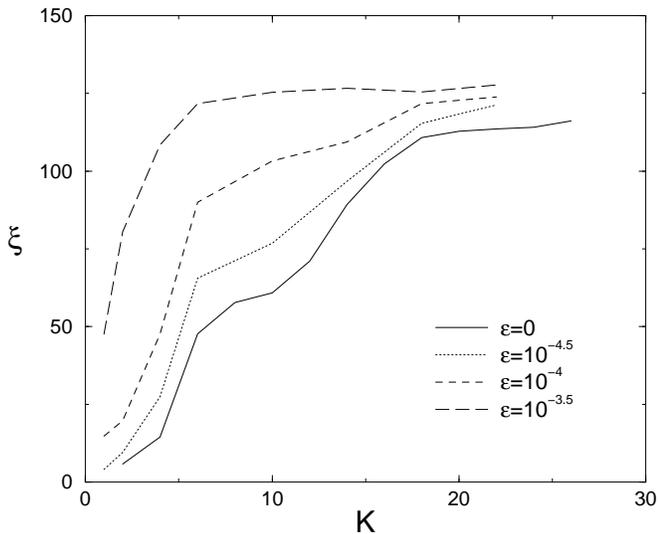}
\caption{\label{transition_L27_nr8} IPR in presence of
imperfections as a function of $K$ in the transition regime.
Here $L=27$, $\hbar/(2\pi)=(13-\sqrt{5})/82$ (actual value is the 
nearest fraction with
denominator $2^{n_r}$), 
initial state is $|\psi_0\rangle=|0\rangle$,
IPR is shown after 100 iterations using $2\times 40$ slices per iteration,
$n_r=8$ ($n_q=n_r+1$), with averages made over
$10$ realizations of disorder.} 
\end{figure}

Fig.\ref{transition_L27_nr8} shows examples of the growth of the IPR 
as a function of $K$ and imperfection strength.  In the
partially delocalized zone, the figure shows a growth of the IPR with $K$,
which is strongly affected by imperfections.
An interesting quantity is the value of the transition point
between localized and delocalized states.  In systems such as the
Anderson model investigated in \cite{pomerans}, the transition point
is well-defined, since all states are localized or delocalized
on one side of the transition.  In the case of the kicked Harper
model there is some arbitrariness
in the definition.  We chose as transition point the value $K_c$
(at $L$ fixed) for which the IPR is $N_H/4=N/8$
(even for a totally delocalized state, the IPR is actually
often $N_H/2=N/4$
instead of $N_H$ due to fluctuations).  
In the partially delocalized regime, the IPR at fixed $K$ should grow
with $\varepsilon$.  If the system is in the Breit-Wigner (golden-rule)
regime, IPR should grow as $\Gamma /\Delta_n$ where $\Delta_n\sim 1/N$
is the mean level spacing and 
$\Gamma\approx 2\pi  |V_{typ}|^2/\Delta_c \sim 
\varepsilon^2 n_g^2 n_q $.
We therefore expect the transition point to move with imperfections
as $\Gamma /\Delta_n\sim \varepsilon^2 n_g^2 n_q N$.  On the contrary,
in the gaussian regime, the IPR grows like $\sigma/\Delta_n$,
where $\sigma\sim \varepsilon n_g\sqrt{n_q}$.  In this case the
transition point should move as $\varepsilon n_g \sqrt{n_q} N$.

\begin{figure}
\includegraphics[width=\columnwidth]{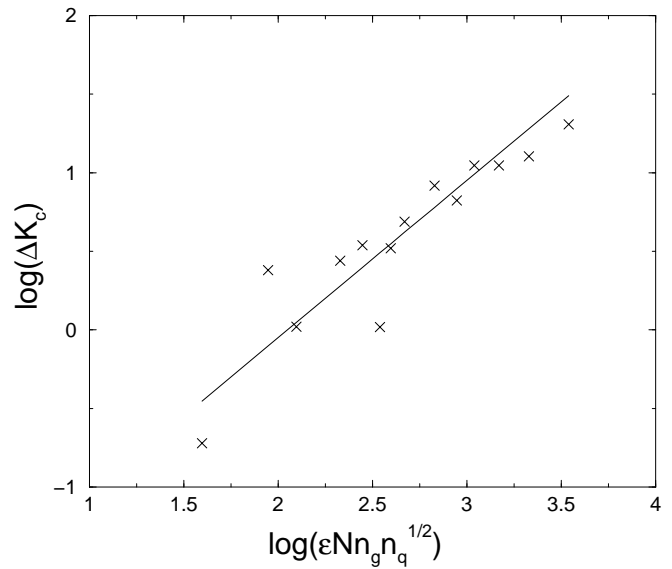}
\caption{\label{transition_scaling} Shift of the transition point due to
imperfections as a function of imperfection strength and $n_q$.
Parameters values are the same as in Fig.\ref{transition_L27_nr8},
with averages made over up to $10$ realizations of disorder.
Solid line corresponds to the dependence
$\Delta K_c\propto \varepsilon n_g \sqrt{n_q} N$. Logarithms are decimal.} 
\end{figure}

Fig.\ref{transition_scaling} shows the data numerically obtained
for the shift of the transition point due to imperfections.
It indicates that $\Delta K_c\sim \varepsilon n_g \sqrt{n_q} N$
agrees with the data, whereas $\varepsilon^2 n_g^2 n_q  N$
is a much less reasonable scaling variable (data not shown).
The data therefore seem to indicate that in the partially delocalized 
regime as in the localized regime, the IPR grows as 
$\varepsilon n_g \sqrt{n_q} N$, as does the shift
of the transition point. This result is in sharp contrast
with the findings of \cite{pomerans} for the Anderson transition, which was
shown to scale polynomially with the number of qubits.  In our case,
the presence of delocalized state coexisting with localized states
makes the delocalization much easier in presence of imperfections.

\begin{figure}
\includegraphics[width=\columnwidth]{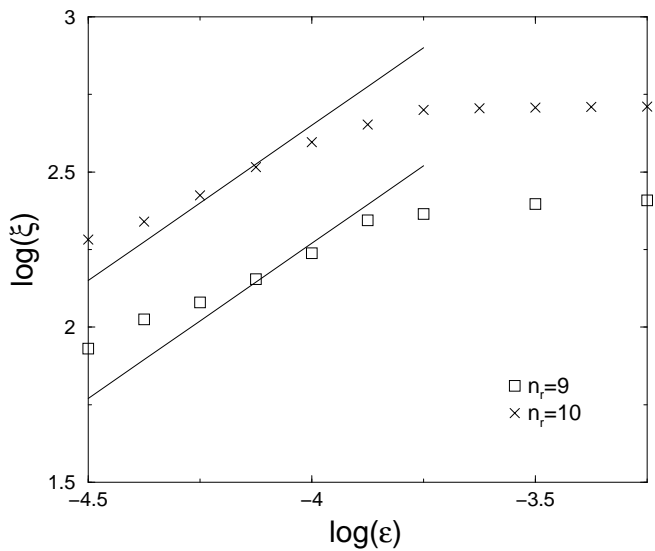}
\caption{\label{IPR_K10} IPR as a function of 
$\varepsilon$  for $K=10$, $L=27$ with
other parameter values the same as in Fig.\ref{psiK2}.
Solid lines correspond to the dependence $\xi \propto \varepsilon$ . 
Logarithms are decimal.} 
\end{figure}

Fig.\ref{IPR_K10} shows the scaling of the IPR as a function of $\varepsilon$.
For small values of $n_q$, the IPR without imperfections is already a large
fraction of Hilbert space dimension, so data are meaningful only for 
$n_r\geq 9$.  Still, data shown on Fig.\ref{IPR_K10} seem to indicate that the
regime where $\xi \propto \varepsilon$ is present, confirming that the 
system is in a gaussian regime rather than in the golden rule regime. 

The scaling laws obtained in this regime show that there is
an exponentially small value $\varepsilon_c$ given by (\ref{thresholdK10})
above which imperfections destroy the localization properties
of the system.  In particular, the transition point 
is exponentially sensitive to the number of qubits.  
This sharp difference between localized and delocalized regime
can be easily seen on experiments: the long time behavior of the system
will be very different in both cases.
Still, the algorithms presented can be useful in
delocalized regime in presence of imperfections, 
even for $\varepsilon>\varepsilon_c$.  Indeed,  the system should remain close
to the exact one up to a time 
$\sim 1/\sigma\sim 1/(\varepsilon n_g \sqrt{n_q})$
as in the localized regime, so measurability of physical
quantities will eventually rest on the comparison of this time scale with
the time for the system to show the delocalization plateau.  On the
contrary, in the localized regime for moderate levels of imperfections
the localization length can be measured for very long times.

\section{Spectrum: measurement and imperfection effects}

\begin{figure}
\includegraphics[width=\columnwidth]{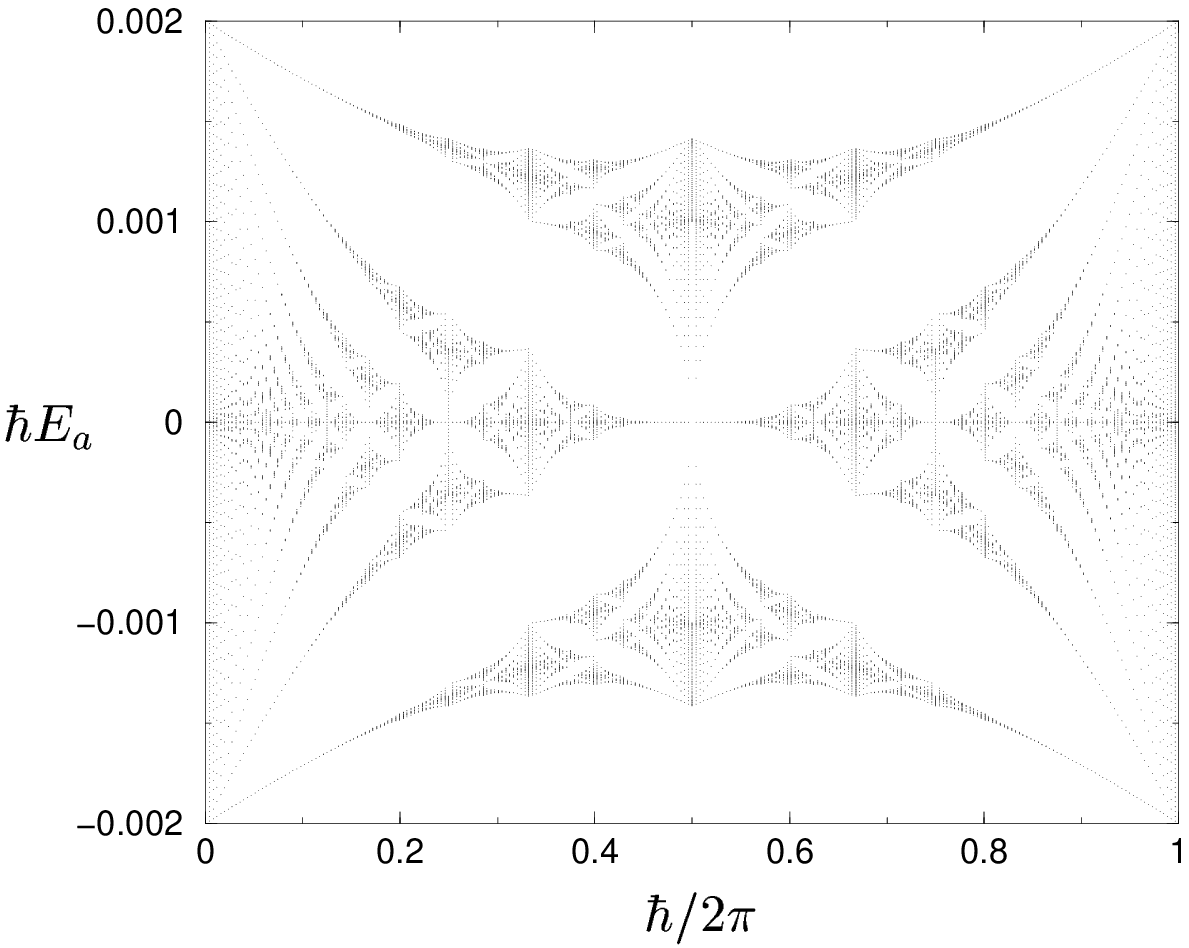}
\caption{\label{papillon8} Eigenphases of the Harper operator (\ref{qharper})
as a function of $\hbar$ for $K=L=10^{-3}$; $n_r=8$.} 
\end{figure}

Another type of physical properties which can be obtained
through quantum simulation of the kicked Harper model 
concerns the spectrum of the evolution operator
$\hat{U}$.  This spectrum has been the focus of many studies 
(see e.g.\cite{artuso}):
it shows multifractal properties, and 
transport properties (localized or delocalized states) are reflected in the
eigenvalues, as well as dynamical properties (chaotic or integrable states). 
Additionally, for small $K=L$, this spectrum 
will be close to the famous spectrum of the Harper model
(``Hofstadter butterfly''), which shows fractal
properties \cite{hofstadter}, as can be seen in Fig.\ref{papillon8}.

To get information about eigenvalues,
we can use the phase estimation algorithm.
This algorithm, at the heart of the Shor algorithm, proceeds
by transforming the state $\sum_t |t\rangle|\psi_0\rangle$
into $\sum_t |t\rangle |U^t \psi_0\rangle$. Then a QFT of the 
first register will give peaks at the values of the eigenphases
of $U$.  To be efficient, this process should be applied to
operators $U$ for which exponentially large iterates can be obtained
in polynomial number of operations. In \cite{ablloyd} it was suggested
that one can obtain approximate eigenvalues exponentially fast
provided one starts from an initial state $|\psi_0\rangle$ already
close to an eigenvector. In the case at hand, we do not know 
how to get exponentially large iterates in polynomial time, nor
how to build a good approximation of the eigenvectors without
knowing them.  We therefore suggest a third strategy, which 
is more generally applicable than the ones above, but does
not yield an exponential gain.  

\begin{figure}
\includegraphics[width=\columnwidth]{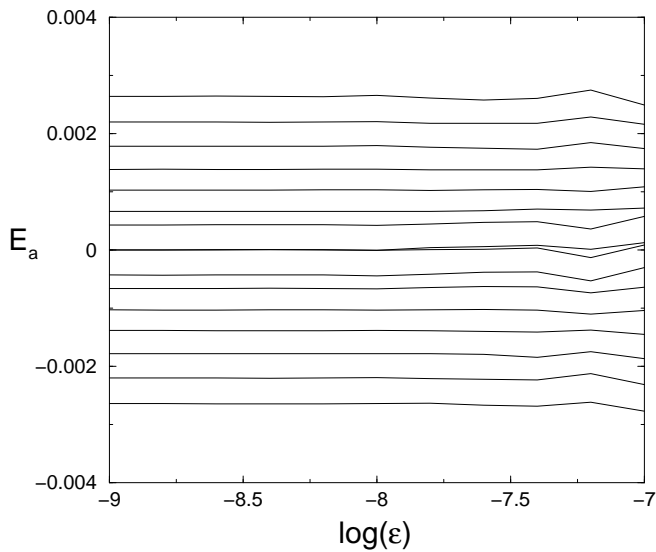}
\caption{\label{slices_spectre_erreurs} Eigenphases of the evolution operator
$\hat{U}$ of (\ref{qharper})
as a function of imperfection strength. The slice method is used
with $2\times 100$ slices to compute the operator.
The $16$ eigenphases
closest to $0$ are shown. Here $n_r=6$ ($n_q=n_r+1$), $\hbar=2\pi/2^6$
(actual value is the nearest fraction with denominator $2^6$), 
$K=L=10^{-3}$. An overall phase factor (global motion of eigenvalues)
has been eliminated.  Logarithm is decimal.} 
\end{figure}

\begin{figure}
\includegraphics[width=\columnwidth]{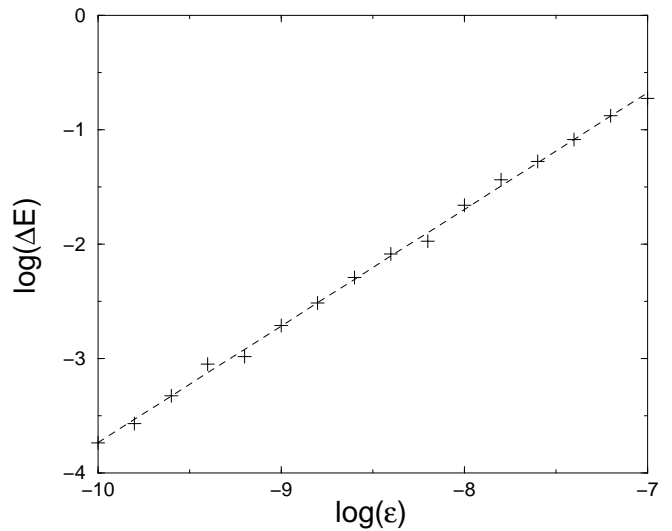}
\caption{\label{slices_spectre_scaling} 
Average error (in units of mean level spacing) of computed eigenphases 
through the slice method as
a function of imperfection strength; parameters are the same as in 
Fig.\ref{slices_spectre_erreurs}, and averages were made over all
eigenvalues and 
over $10$ realizations of disorder. Dashed line corresponds
to $\Delta E \propto \varepsilon$.  Logarithms are decimal.}
\end{figure}

We first build the state $\sum_{t=0}^{N_H-1} |t\rangle|\psi_0\rangle$,
where $|\psi_0\rangle$ is an arbitrary quantum state 
on a Hilbert space of dimension $N_H=2^{n_r}$
which can be efficiently built,
for example it can be the state $2^{-n_r/2}\sum_n |n\rangle$,
which can be obtained from $|0\rangle$ with $n_r$ Hadamard gates.
Once the state 
$|0\rangle|\psi_0\rangle$ is realized, it can be
transformed with $n_r$ Hadamard gates on the first register into
$2^{-n_r/2}\sum_{t=0}^{N_H-1} |t\rangle|\psi_0\rangle$.  We have seen that 
the evolution operator $U$ can be implemented in poly(log$N_H$) operations
by the three strategies exposed in section III.  Therefore we can apply
powers of $U$ on the second register controlled by the value of the
first register.  This yields $2^{-n_r/2}\sum_t |t\rangle|U^t\psi_0\rangle$
in $O(N_H)$ operations, up to logarithmic factors.  A QFT of the first register
will yield peaks centered at eigenvalues of the operator $U$.  Thus
measurement of the first register will give an eigenvalue of $U$ with good
probability in $O(N_H)$ operations including measurement.  A drawback of this 
approach is that peaks have additional probabilities on nearby locations, 
and since the number of eigenvalues is $N_H$, measuring the precise shape
of all peaks will be inefficient ($O(N_H^2)$ operations).  A more precise,
although slower method is to use amplitude amplification \cite{brassard}
(a method derived from the Grover algorithm \cite{grover})
to zoom on a small part of the spectrum.  This enables to get 
the precise values of all eigenvalues in a given interval.  The total
cost will be $O(N_H\sqrt{N_H})$ operations. This methods
which uses Grover's search on phase estimation 
can be seen as a process reverse to quantum counting
\cite{counting} (where phase estimation is used on the Grover
operator).

Calculating the spectrum by direct diagonalization of a $N_H \times N_H$ matrix
such as the one of the operator $\hat{U}$ of (\ref{qharper})
requires in general of the order of $N_H^3$ classical operations.  
However, in the case of the
operator $\hat{U}$ of (\ref{qharper})
there is a faster classical method similar to the quantum phase
estimation algorithm: one iteration of $\hat{U}$ can be computed classically in
$O(N_H)$ operations (up to logarithmic factors) by using the classical FFT
algorithm to shift between $n$ and $\theta$ representations, and multiplying
by the relevant phase in each representation.  Iterating this process
$N_H$ times and keeping each intermediate wave function costs 
$O(N_H^2)$ operations.  Then a FFT enables to get the spectrum of $U$ 
with $O(N_H\log N_H)$ operations.  
This method was advocated in \cite{ketzmerick}
for getting the spectrum of the kicked Harper model.  Therefore it is
possible to get the spectrum classically in $O(N_H^2)$ operations up
 to logarithmic factors.  Thus the quantum algorithms explained above 
($O(N_H)$ operations for one eigenvalue with unknown precision, 
$O(N_H\sqrt{N_H})$ for all eigenvalues in a given small interval) realize
a polynomial gain compared to the classical ones.  It is important
to note that although the number of operations needed is only 
polynomially better in the quantum case, the spatial resources
are exponentially smaller (logarithmic number of qubits compared to 
the number of classical bits).  

\begin{figure}
\includegraphics[width=\columnwidth]{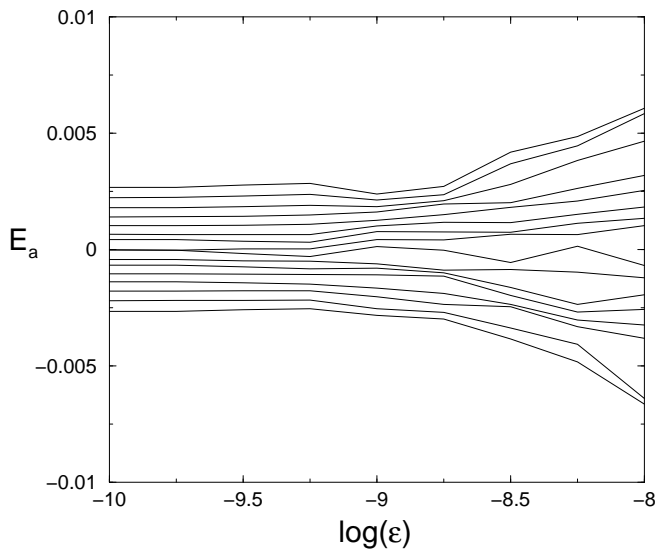}
\caption{\label{chebyshev_spectre_erreurs} Eigenphases of the 
evolution operator
$\hat{U}$ of (\ref{qharper})
as a function of imperfection strength. The Chebyshev method is used; a
Chebyshev polynomial of degree $6$ is taken, keeping all gates.
The $16$ eigenphases
closest to $0$ are shown. Here $n_r=6$ ($n_q=n_r$), $\hbar=2\pi/2^6$
(actual value is the nearest fraction with denominator $2^6$), 
$K=L=10^{-3}$. An overall phase factor (global motion of eigenvalues)
has been eliminated. Logarithm is decimal.} 
\end{figure}

\begin{figure}
\includegraphics[width=\columnwidth]{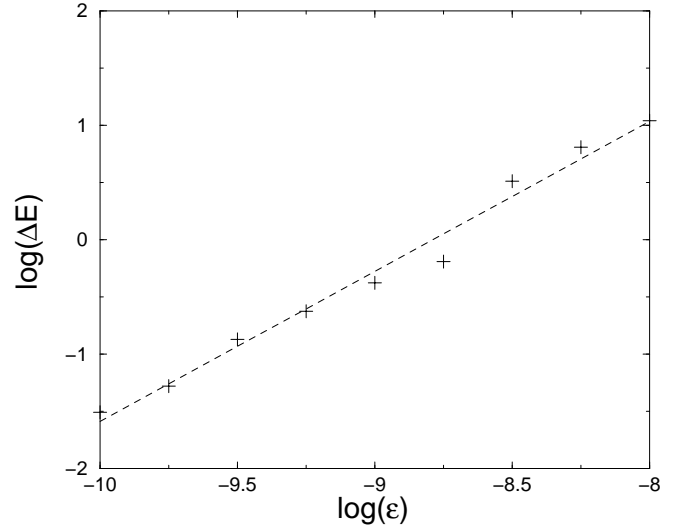}
\caption{\label{chebyshev_spectre_scaling} 
Average error (in units of mean level spacing) of computed eigenphases
through the Chebyshev method as
a function of imperfection strength; parameters are the same as in 
Fig.\ref{chebyshev_spectre_erreurs}, and averages were made over all
eigenvalues. Dashed line corresponds
to $\Delta E \propto \varepsilon^{1.3}$.  Logarithms are decimal.} 
\end{figure}

The Fig.\ref{slices_spectre_erreurs}-\ref{chebyshev_spectre_erreurs}
show the spectrum of the kicked Harper model in
presence of errors for both slice and Chebyshev methods.
The error model chosen is the static imperfection Hamiltonian
(\ref{hamil}) as in the preceding section.
The evolution operator was computed
by evolving basis states
in presence of errors and then diagonalizing the resulting
operator.  The spectrum shown corresponds to small $K=L$, where
the spectrum is close to the ``Hofstadter butterfly'', as can be seen
in Fig.\ref{papillon8}.
Only $16$ eigenvalues are shown.  Overall phase shifts due to errors
were eliminated since it seems reasonable they can be estimated and
compensated.
It is clear from the data
presented that eigenvalues are much more sensitive to strength of errors
than transport properties.  Numerical limitations
prevented us to find the scaling in $n_q$ of error effects, 
but Fig.\ref{slices_spectre_scaling}-\ref{chebyshev_spectre_scaling} 
show the scaling with respect to $\varepsilon$ at constant $n_q$.

In the case of the slice method, the average error on the eigenvalue
is clearly linear in $\varepsilon$.  We think
this corresponds probably
to a perturbative regime, since small values of $\varepsilon$ are involved.
For the Chebyshev method, 
our data indicate that a lower level of errors is needed than in the 
slice method to get good accuracy.  This could have been expected, since
we established in Section III than this method necessitates more gates
for a similar accuracy, and each gate introduces errors.  The scaling of errors
with respect to $\varepsilon$ indicates the law 
$\Delta E \sim \varepsilon^\alpha$ 
with $\alpha\approx 1.3$. 

\section{Conclusion}

In this paper, several quantum algorithms were presented
which enable to simulate the quantum kicked Harper model,
a complex system with relevance to certain physical problems.
The comparison showed that while the slice method and the Chebyshev method 
are approximate, they are much more economical in resources than the
exact simulation.  It was also shown that different transport and spectral
properties can be obtained more efficiently on a quantum computer than
classically, although the gain is only polynomial.  Numerical
simulations enabled us to precise the effect of numerical errors 
on these algorithms, and also to evaluate the effects of imperfections.
The results show that depending on the regime of parameters,
the same quantity can be stable or exponentially sensitive to imperfections.
In general, in presence of moderate amount of errors the
results of the algorithm can be meaningful, but a careful choice of 
the measured quantities should be done.
  For
the different quantities computed, the slice method was shown to be
more efficient and resilient to errors than the Chebyshev method, although
the latter is similar to the method used in classical computers to 
evaluate functions.

Our results show that interesting quantum effects such that
fractal-like spectrum, localization properties, anomalous diffusion
are already visible with 7-8 qubits.
We therefore believe that such algorithms could be used in experimental
implementations in the near future.

We warmly thank Dima Shepelyansky for many helpful suggestions
in the course of this work, and also Andrei Pomeransky and
Klaus Frahm
for several discussions.
We thank the IDRIS in Orsay and CalMiP in Toulouse for 
access to their supercomputers.
This work was supported in part by the NSA
and ARDA under ARO contract No. DAAD19-01-1-0553, 
by the EC  RTN contract HPRN-CT-2000-0156 and by the 
project EDIQIP of the IST-FET program of the EC.

\end{document}